\documentclass[]{partc}
\usepackage{graphicx}

\usepackage[hidelinks]{hyperref}

\usepackage{comment}
\usepackage{tabularx}
\usepackage{subcaption}
\usepackage{multirow}
\usepackage{xcolor}
\usepackage{amsfonts}
\usepackage{bm}
\usepackage{makecell}
\usepackage{array}

\title{Characterizing Lane-Changing Behavior in Mixed Traffic}

\author[1]{Sungyong Chung\orcidlink{0009-0001-3131-6011}}

\author[1]{Alireza Talebpour\orcidlink{0000-0002-5412-5592}\thanks{Corresponding author. Email: ataleb@illinois.edu}}

\author[2]{Samer H. Hamdar\orcidlink{0000-0001-6896-367X}}

\affil[1]{Department of Civil and Environmental Engineering, University of Illinois at Urbana-Champaign}
\affil[2]{Department of Civil and Environmental Engineering, George Washington University}

\date{}

\begin{document}
\maketitle
\begin{abstract}
Characterizing and understanding lane-changing behavior in the presence of automated vehicles (AVs) is crucial to ensuring safety and efficiency in mixed traffic. Accordingly, this study aims to characterize the interactions between the lane-changing vehicle (active vehicle) and the vehicle directly impacted by the maneuver in the target lane (passive vehicle). Utilizing real-world trajectory data from the Waymo Open Motion Dataset (WOMD), this study explores patterns in lane-changing behavior and provides insight into how these behaviors evolve under different AV market penetration rates (MPRs). In particular, we propose a game-theoretic framework to analyze cooperative and defective behaviors in mixed traffic, applied to the 7,636 observed lane-changing events in the WOMD. First, we utilize k-means clustering to classify vehicles as cooperative or defective, revealing that the proportions of cooperative AVs are higher than those of HDVs in both active and passive roles. Next, we jointly estimate the utilities of active and passive vehicles to model their behaviors using the quantal response equilibrium framework. Empirical payoff tables are then constructed based on these utilities. Using these payoffs, we analyze the presence of social dilemmas and examine the evolution of cooperative behaviors using evolutionary game theory. Our results reveal the presence of social dilemmas in approximately 4\% and 11\% of lane-changing events for active and passive vehicles, respectively, with most classified as Stag Hunt or Prisoner's Dilemma (Chicken Game rarely observed). Moreover, the Monte Carlo simulation results show that repeated lane-changing interactions consistently lead to increased cooperative behavior over time, regardless of the AV penetration rate.
\end{abstract}


\section{1. Introduction}
As automated vehicles (AVs) become increasingly integrated into human-dominated traffic systems, characterizing their interactions with human-driven vehicles (HDVs) is essential for improving the safety, efficiency, and stability of mixed traffic flows. These interactions are particularly complex during lane-changing (LC) maneuvers, where strategic decision-making and real-time assessment of other drivers' intentions are essential. While much of the existing research has focused on the broader impacts of AVs on traffic flow through simulations using assumed AV models \cite{Talebpour2016,Zheng2020} or calibrated models with empirical trajectory data \cite{Shang2021,Shi2021,Chung2024}, studies on how AVs affect human driver behavior are relatively rare, and those specifically examining their influence during lane changes are even more scarce. Understanding these behavioral shifts is vital for improving traffic safety and efficiency. Moreover, analyzing how these dynamic interactions evolve will provide valuable insights for managing mixed traffic flows as AVs become more prevalent in future transportation systems.

A few studies have explored how the behavior of HDVs shifts with the introduction of AVs. \citep{Rahmati2019} conducted an experimental study to examine whether human-human interactions on the road differ from human-AV interactions. Their findings revealed a statistically significant difference in human driving behavior between the two scenarios and emphasized the importance of incorporating these behavioral changes into microscopic simulation models to better represent mixed traffic. Using the same dataset, \citep{Zhang2024} found that the Intelligent Driver Model (IDM) fails to capture human behavior changes when interacting with AVs. They applied the stochastic IDM model, showing that human driver uncertainty decreases when following an AV. Similarly, \citep{Zhao2020} carried out field experiments to investigate differences in car-following behavior when human drivers follow an AV compared to an HDV across various scenarios. Their findings indicated that human drivers responded differently to AVs only when they were visually identifiable. \citep{Mahdinia2021} conducted a field experiment with nine drivers under two scenarios: following an AV using a human driver speed profile or an AV using its own speed profile. The results showed that AVs in mixed traffic can influence human driver behavior, leading to more stable traffic flow, lower crash risk, and reduced fuel consumption. \citep{Soni2022} conducted a field test with 18 participants, finding that human drivers accepted smaller gaps and maintained shorter headways when interacting with AVs compared to HDVs, though no significant difference was observed in car-following behavior.

Despite these valuable insights, the studies mentioned above rely on controlled field experiments with limited sample sizes. As \citep{Zhang2023} noted, such results may lack generalizability to naturalistic driving environments. To address this limitation, they used NGSIM \textcolor{black}{\cite{USDOT2016}} and Lyft Level 5 prediction \textcolor{black}{\cite{Houston2021}} datasets to examine how AVs influence human drivers' car-following behavior. Their results showed a significant difference, with human drivers maintaining shorter and less variable time headways when following an AV compared to an HDV. Similarly, \citep{Wen2023} employed inverse reinforcement learning utilizing the Waymo Open Dataset (WOD) \cite{Sun2020Waymo} and the Waymo Open Motion Dataset (WOMD) \cite{Ettinger2021} to analyze the differences between HDV-following-AV and HDV-following-HDV dynamics. The recovered reward functions of the drivers of the following vehicles revealed that drivers’ preferences differ when following AVs compared to HDVs. 

While these studies focus primarily on longitudinal behavior, little attention has been paid to lane-changing dynamics. Accordingly, the main focus of this study is to fill that gap by introducing a systematic approach to examine lane-changing interactions between the lane-changing vehicle and the interacting vehicle in the target lane, using empirical data. In particular, utilizing game theory and WOMD, we illustrate how real-world AV and HDV trajectories can be explored to characterize these complex interactions during lane-changing maneuvers. It is important to note that this study is focused on methodology rather than drawing conclusions based on WOMD, considering the inaccuracies in that dataset \cite{zhang2025can}.

Game theory, known for modeling strategic decision-making in multi-agent settings, has been extensively applied to analyzing lane-changing behavior \cite{Ji2020}. In these studies, the lane-changing vehicle typically faces the decision to either change lanes or maintain its lane, while the lag vehicle can adopt various responses, such as (i) yielding or not yielding \cite{Kita1999,Kita2002}, (ii) accelerating, decelerating, or changing lanes \cite{Talebpour2015}
, or (iii) accelerating or yielding \cite{Yu2018}. Our work shifts the focus to how AVs and HDVs strategically choose to cooperate or defect during lane changes once the maneuver is initiated. In the context of lane-changing, cooperative and defective behaviors can create social dilemmas, situations where cooperation improves overall traffic safety but leaves cooperative drivers vulnerable to exploitation by more aggressive maneuvers, such as forcing a merge. These dynamics reflect classic game-theoretic models like the Stag Hunt and Prisoner's Dilemma, where cooperation depends on mutual trust, but defection remains a constant temptation. \citep{Tanimoto2014} classified drivers into two categories: cooperative agents who always stay in their lane and defective agents who attempt to change lanes to move ahead. By defining the lane-changing reward as the average velocity of each driver type on the road, they demonstrated, through numerical simulations, that the Prisoner's Dilemma arises at moderate traffic densities. However, their work relied on simulations rather than real-world lane-changing data. In this study, following a similar approach to \citep{Leibo2017}, we apply an empirical payoff matrix framework to quantify cooperative and defective behaviors in lane-changing interactions using the real-world WOMD dataset. This approach allows us to identify potential social dilemmas in lane-changing in real-world scenarios.

While the empirical payoff matrix framework helps quantify cooperative and defective behaviors, it does not fully capture the evolving nature of interactions over time. Therefore, we also aim to examine how the behavior of AVs and HDVs evolves under different levels of AV penetration rates on the road. Traditional game theory typically assumes fixed strategies and rational decision-making. In contrast, evolutionary game theory offers a more comprehensive framework by modeling how behaviors change over time through repeated interactions. Only a few studies have applied evolutionary game theory to understand how behaviors evolve in traffic flow. \citep{CortesBerrueco2016} applied evolutionary game theory to study the evolution of cooperation in traffic flow under varying traffic densities. They considered two possible behaviors, i.e., cooperative and defective, and separately defined lane-changing models for each, with cooperative drivers adopting more cautious lane-changing decision parameters. They defined the payoff as the speed gain during lane changes and simulated the evolution using numerical methods. However, their results heavily rely on the assumed lane-changing models for cooperative and defective drivers. More recently, \citep{Rahmati2021} proposed a framework to analyze the evolution of driver behavior in mixed traffic using evolutionary game theory. Their approach, however, was based on assumed payoffs for AVs and HDVs, rather than based on real-world observations. As a result, empirical evidence on how cooperative and defective behaviors in lane-changing interactions evolve in mixed traffic remains unexplored.

In order to address the aforementioned gaps, the main contributions of this study are: (1) Introducing a systematic approach to classify lane-changing behaviors of AVs and HDVs into cooperative and defective using k-means clustering; (2) Introducing a methodology to identify and analyze social dilemmas in lane-changing maneuvers by quantifying the utilities associated with each behavior and utilizing these utilities to construct empirical payoff tables; and (3) Introducing a framework to model the evolution of cooperative and defective behaviors of AVs and HDVs under different AV penetration rates using evolutionary game theory. By combining real-world trajectory data with game-theoretic frameworks, this study lays the foundation for characterizing and understanding lane-changing behaviors in mixed traffic. 

The remainder of this paper is organized as follows: Section 2 presents the dataset used in this study and the procedure for extracting lane-changing events; Section 3 introduces the proposed framework for analyzing social dilemmas and the evolution of cooperative behavior in lane-changing; Section 4 provides the empirical results and discusses key findings; and Section 5 concludes the paper.

\section{2. Data Description}
This section provides an overview of the WOMD and outlines how lane-changing trajectories were extracted from the dataset.

\subsection{2.1. Waymo Open Motion Dataset}
The WOMD \textcolor{black}{\cite{Ettinger2021}} was utilized to examine lane-changing behaviors in diverse traffic scenarios. The WOMD includes data from various locations, such as San Francisco, Mountain View, Los Angeles, Detroit, Seattle, and Phoenix. Each scenario in the dataset provides detailed trajectory data for Waymo vehicles and other objects detected within the range of the onboard sensors. These data capture a wide range of lane-changing events, including interactions between Waymo vehicles and HDVs, as well as interactions exclusively involving HDVs. 

The WOMD consists of 20-second trajectory segments sampled at 10 Hz. The dataset includes information such as scenario IDs, unique object tracking IDs, object types (e.g., vehicles, pedestrians, and cyclists), Waymo vehicle identification, and detailed object attributes like position (x, y, z), dimensions (length, width, height), heading, and velocity. Additionally, it provides map features for each scenario such as lane centers, lane boundaries, and road boundaries represented as 3D polylines. For more details about this dataset, refer to \citep{Ettinger2021}.

\subsection{2.2. Lane-Changing Events Extraction}
To identify and analyze lane-changing events, we implemented a systematic three-step process. First, vehicles were assigned to lanes by determining the nearest point along the lane center to each vehicle's position. As vehicles moved, their lane center assignment was updated to the closest point from either the current lane center ID, any exit lane center IDs, or the adjacent left or right lane center IDs, as specified in the map features. To prevent vehicles from being assigned to an incorrect lane center ID, such as a parked vehicle that shouldn't be linked to any lane, a threshold of 3.5 meters was applied for the distance to the nearest lane center point.

Next, lane-changing events were detected by identifying instances where a vehicle transitioned from its current lane center ID to an adjacent left or right lane center ID. Events within intersections were omitted from the analysis by filtering out interpolated lanes, which represent virtual connections at intersections. Additionally, to exclude lane changes associated with turnings (e.g., right or left turns), only events where the difference in heading angle before and after the lane change was less than 0.2 radians were included.

\begin{figure}[tb!]
    \centering
    \includegraphics[width=\textwidth]{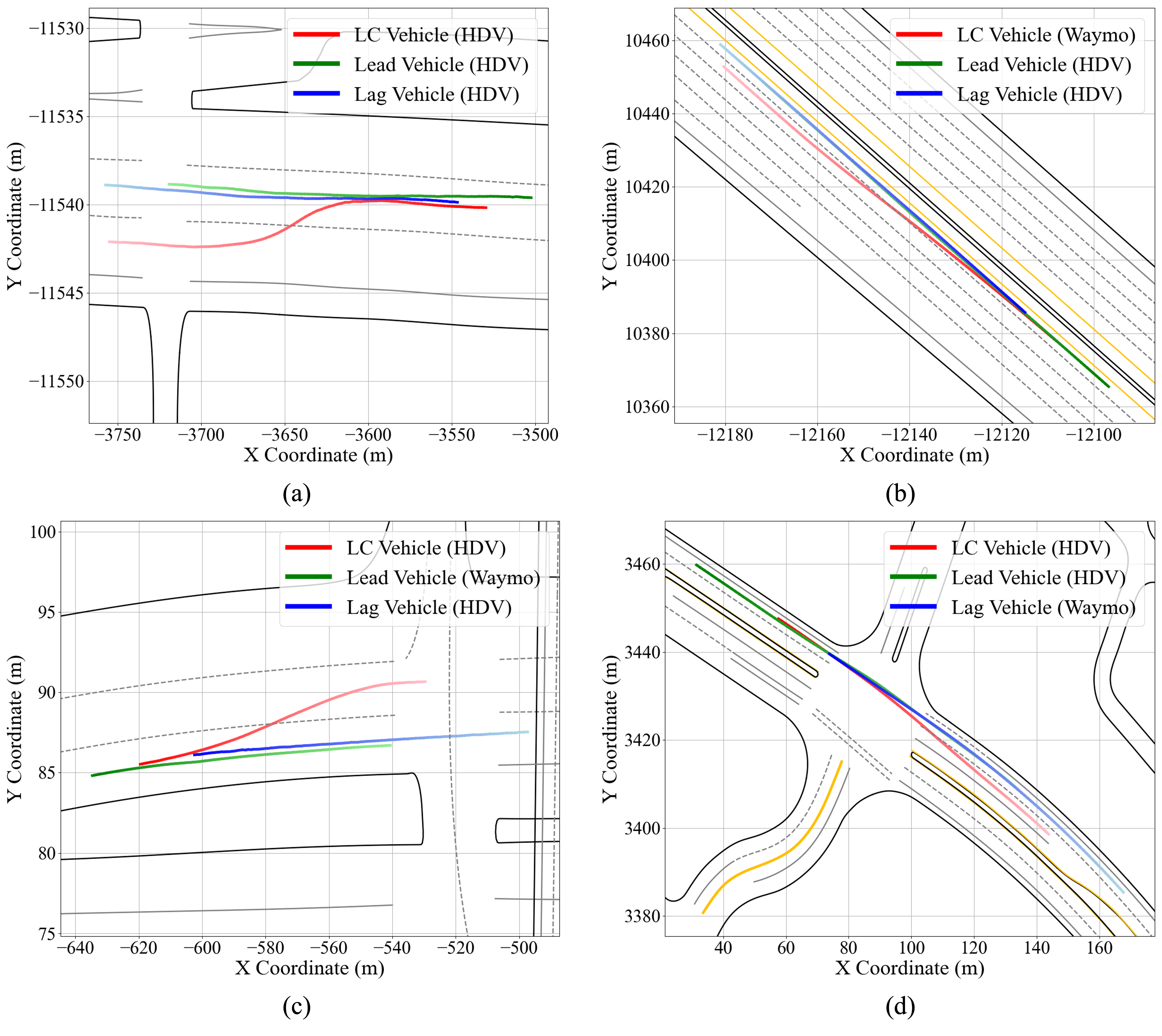}
    \caption{Sample \textcolor{black}{10-second} trajectories of extracted lane-changing (LC) events: (a) HDV only, (b) Waymo as a LC vehicle, (c) Waymo as a lead vehicle, and (d) Waymo as a lag vehicle. \textcolor{black}{The starting points of all vehicles are shown in lighter colors, and the ending points in darker colors. The LC vehicle is shown in red, the lead vehicle in green, and the lag vehicle in blue.} The road edge is shown as a black solid line, white solid and dashed lane markings on the road are represented as gray solid and dashed lines in the figure, and the yellow centerline is depicted as a yellow solid line.}
    \label{fig:lane_changing_trajectories}
\end{figure}

Finally, at each time step, each vehicle was assigned a leader and a follower to identify the lead and lag vehicles involved in lane-changing events. Figures \ref{fig:lane_changing_trajectories}(a-d) illustrate the 10-second trajectories of lead, lag, and lane-changing vehicle from four distinct lane-changing event scenarios: (a) involving only HDVs, (b) with Waymo as the lane-changing vehicle, (c) with Waymo as the lead vehicle, and (d) with Waymo as the lag vehicle. Note that we focus only on lane-changing events in which information about both lead and lag vehicles is available. Furthermore, we selected lane-changing events with an average vehicle speed below 25 $m/s$ to focus on reasonable speeds for urban roads. After applying these criteria, a total of 7,636 lane-changing events were extracted for analysis.

\begin{figure}[tb!]
    \centering
    \begin{subfigure}[t]{0.7\textwidth}
        \centering
        \includegraphics[width=\textwidth]{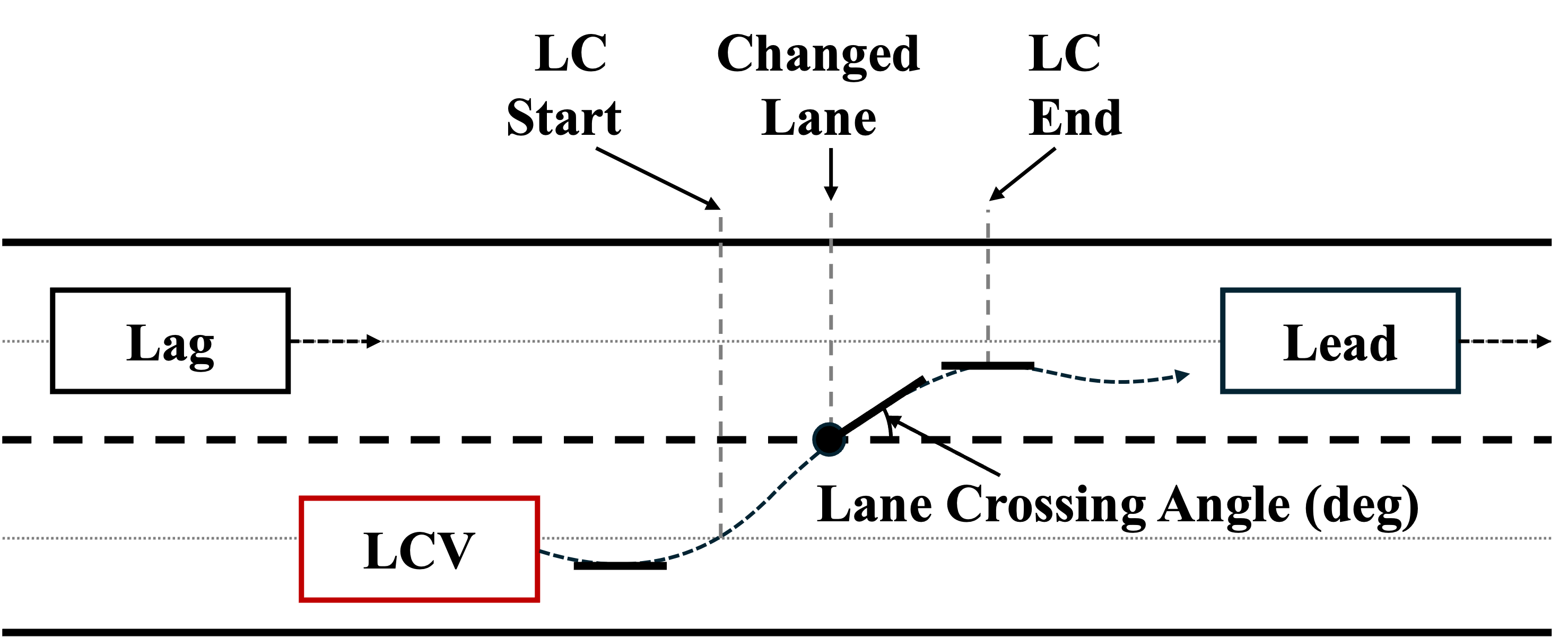}
        \caption*{(a)}
        \label{fig:defining_variables_1}
    \end{subfigure}
    
    \vspace{0.5cm} 
    
    \begin{subfigure}[t]{0.7\textwidth}
        \centering
        \includegraphics[width=\textwidth]{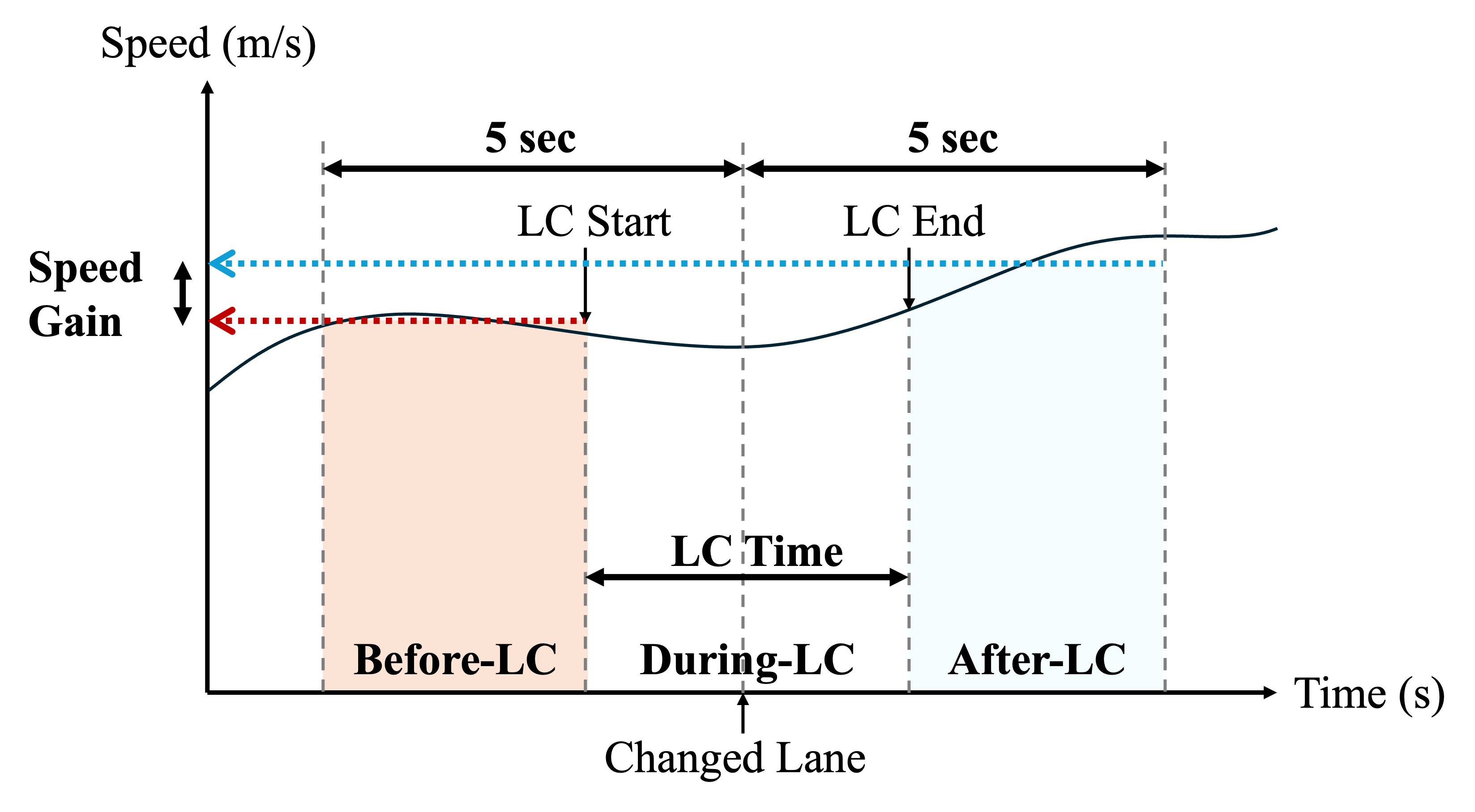}
        \caption*{(b)}
        \label{fig:defining_variables_2}
    \end{subfigure}
    
    \caption{Illustration of defined variables from the Waymo Open Dataset. (a) Definition of lane-changing timing, including the start, lane change, and end points, as well as the lane crossing angle. (b) Definition of speed gain and the time periods before, during, and after the lane change.}
    \label{fig:defining_variables}
\end{figure}

To perform a comprehensive analysis of lane-changing events, we defined key variables to characterize these events. As illustrated in Figure \ref{fig:defining_variables}(a), we first identified the start, end, and timing of the lane change. The timing of the lane change was determined by tracking the vehicle's closest lane center ID, as described previously. When the closest lane center ID shifted from its current lane to an adjacent lane, this moment was recorded as the lane-changing timing. Next, we examined the distance of the vehicle from the lane center at each time step before and after the lane change. By identifying the point where this distance continuously decreased and then increased, we marked the point as the start and end points of the lane change. This can occur either when a vehicle does not reach the center of the target lane (as shown by the LC End point in Figure \ref{fig:defining_variables}(a)) or when a vehicle crosses the lane center (as indicated by the LC Start point in Figure \ref{fig:defining_variables}(a)). Additionally, the lane-crossing angle was defined as the angle between the lane center and the vehicle's moving direction at the time of the lane change.

Each lane-changing event was divided into three distinct periods, as illustrated in Figure \ref{fig:defining_variables}(b). For each event, we extracted the trajectories of the lead, lag, and lane-changing vehicles from 5 seconds before the lane change to 5 seconds after the lane change, resulting in a total duration of 10 seconds. Using the identified start and end timings of the lane change, we segmented this duration into the following periods: the before-, during-, and after-lane-changing periods. The duration of the "during-lane-changing" period was defined as the lane-changing time in this study. Lastly, we defined the speed gain as the average speed during the "after-lane-changing" period minus the average speed during the "before-lane-changing" period.

The key variables derived from the lane-changing trajectories, along with other straightforward metrics such as speed, acceleration, and yaw rate, are utilized in the following section to classify the behaviors of both the lane-changing vehicle and the lag vehicle, and to systematically quantify the characteristics of lane-changing events.

\section{3. Methodology}
This section is organized into two subsections. The first subsection presents a framework that explores the existence of social dilemmas in lane changes. The second subsection introduces a framework that applies Monte Carlo simulations to examine the evolution of cooperation in lane-changing behavior within the mixed traffic of AVs and HDVs.

\subsection{3.1. Social Dilemma in Lane-Changing}
To investigate the presence of social dilemmas in lane-changing behaviors, we propose an empirical game-theoretic analysis to construct empirical payoff tables that capture the outcomes of one-shot decisions between cooperative and defective behaviors. As previously mentioned, we define the lane-changing vehicle (hereafter referred to as the active vehicle) and its lag vehicle (hereafter referred to as the passive vehicle) as the interacting players in the context of lane-changing. The following four steps are analyzed in this study to identify social dilemmas in lane-changing maneuvers.

\subsubsection{3.1.1. Step 1: Clustering Lane-Changing Behaviors}
In this step, we cluster the lane-changing behaviors of active and passive vehicles into two groups: cooperative and defective. The determination of the optimal number of clusters is a crucial step in the clustering process and can be based on clustering quality metrics or the subjective judgment of researchers \cite{Mohammadnazar2021}. For this study, we choose two clusters to classify drivers in active and passive roles based on their aggressiveness, labeling them as either cooperative or defective. Although we refer to the two clusters as cooperative and defective throughout the study, we acknowledge that this terminology is open to interpretation. \textcolor{black}{Moreover, the selection of two clusters and the corresponding labels} were chosen primarily to align with established conventions in the literature on social dilemmas and evolutionary game theory. As will be demonstrated in the following sections, the classification of social dilemmas is based on the underlying game structures rather than the naming convention. 

Numerous studies have aimed to classify drivers by their level of aggressiveness, but a clear definition of aggressiveness is still lacking in the literature. Several studies define aggressiveness by the frequency of specific actions, such as failing to yield the right of way, disobeying traffic signals, or driving too fast for road conditions \cite{Sun2012}, while others consider behaviors like frequent lane changes \cite{Li2015}. Another approach involves field experiments in which drivers are asked to drive aggressively or conservatively based on their own interpretation of aggressiveness. One study in this category found that driver aggressiveness could be classified using driving behavior, including speed, longitudinal acceleration, and lateral acceleration \cite{Gonzalez2014}. 

However, relatively few studies have specifically analyzed driver aggressiveness during lane-changing maneuvers. For instance, using the empirical dataset, \citep{Das2022} and \citep{Qian2024} applied k-means clustering to categorize lane-changing drivers into two and three groups, respectively, based on their aggressiveness. These studies defined features related to the active vehicle to characterize driver behavior. While prior research has focused solely on the behavior of the active vehicle, we extend their work by classifying the behaviors of both the active vehicle and the passive vehicle during lane-changing interactions. Our analysis incorporates a broader set of features derived from the high-resolution WOMD trajectories, providing a more comprehensive understanding of driver behavior in these interactions. 

Table \ref{tab:behavioral_features_combined} shows the selected behavioral features for active and passive vehicles. These features are utilized to characterize the lane-changing behavior. For instance, for the active vehicle, a longer lane-changing time, denoted as $x^{active}_1$, can potentially indicate cooperative behavior. Similarly, a smaller lane crossing angle, denoted as $x^{active}_5$, can also be indicative of cooperative behavior. However, it is also possible for an active driver to exhibit both a long lane-changing time and a large lane crossing angle. Accordingly, to capture the diverse behaviors in lane-changing maneuvers, we first define 10 active vehicle-related features. Then, we apply k-means clustering to divide the active drivers into two groups based on these feature values. Finally, we label the groups as cooperative and defective, respectively, based on the descriptive statistics of these 10 features within each group. Similarly, for the passive vehicle, we define 4 features related to its behavior during lane change  and apply the k-means clustering to divide the passive vehicles into cooperative and defective.

\begin{table}[tb!]
    \centering
    \caption{Behavioral features of active and passive vehicles.}
    \label{tab:behavioral_features_combined}
    \begin{tabular}{ll}
        \hline
        \textbf{Feature} & \textbf{Description} \\
        \hline
        \multicolumn{2}{l}{\textit{\textbf{Active Vehicle}}} \\
        $x^{active}_1$ & Lane-changing time ($s$) \\
        $x^{active}_2$ & Standard deviation of speed during LC ($m/s$) \\
        $x^{active}_3$ & Speed gain ($m/s$) \\
        $x^{active}_4$ & Maximum heading difference during LC ($rad$) \\
        $x^{active}_5$ & Lane crossing angle ($deg$) \\
        $x^{active}_6$ & Standard deviation of yaw rate during LC ($rad/s$) \\
        $x^{active}_7$ & Maximum lateral speed during LC ($m/s$) \\
        $x^{active}_8$ & Maximum lateral acceleration during LC ($m/s^2$) \\
        $x^{active}_9$ & Standard deviation of acceleration during LC ($m/s^2$) \\
        $x^{active}_{10}$ & Maximum acceleration during LC ($m/s^2$) \\
        \hline
        \multicolumn{2}{l}{\textit{\textbf{Passive Vehicle}}} \\
        $x^{passive}_1$ & Speed gain ($m/s$) \\
        $x^{passive}_2$ & Maximum acceleration during LC ($m/s^2$) \\
        $x^{passive}_3$ & Minimum acceleration during LC ($m/s^2$)\\
        $x^{passive}_4$ & Standard deviation of speed during LC ($m/s$) \\
        \hline
    \end{tabular}
\end{table}

\subsubsection{3.1.2. Step 2: Joint Estimation of Utilities}
In this step, our goal is to predict the behaviors during the lane-changing based on the observations before the lane-changing instance. In other words, we aim to characterize the behaviors of of both active and passive vehicles during the "during-lane-changing" period, based on the state variables defined in Table~\ref{tab:lane_changing_variables}, which are measured during the "before-lane-changing" period. Recall that Figure~\ref{fig:defining_variables}(b) illustrates how these two periods are defined. This approach is based on the core assumption that both active and passive vehicles decide whether to cooperate or defect in a lane-changing event based on the state they observe just prior to the lane change. To mitigate multicollinearity among the state variables, especially with the active vehicle’s average speed ($s_1$), we use the relative average speeds of the lead and passive vehicles ($s_5$ and $s_9$, respectively) instead of their absolute average speeds. We define state vector for the observed lane-changing event $i$, denoted as \( \mathbf{S}_i = \{{s}_1, {s}_2, \dots, {s}_{11}\}_i \).

\begin{table}[tb!]
    \centering
    \caption{Lane-changing state variables.}
    \begin{tabular}{ll}
        \hline
        \textbf{Variable} & \textbf{Description} \\
        \hline
        $s_1$ & Active vehicle average speed before LC ($m/s$) \\
        $s_2$ & Active vehicle standard deviation of speed before LC ($m/s$) \\
        $s_3$ & Active Vehicle average acceleration before LC ($m/s^2$) \\
        $s_4$ & Average lead gap before LC ($s$) \\
        $s_5$ & Lead vehicle relative average speed before LC ($m/s$) \\
        $s_6$ & Lead vehicle standard deviation of speed before LC ($m/s$) \\
        $s_7$ & Lead vehicle average acceleration before LC ($m/s^2$) \\
        $s_8$ & Average lag gap before LC ($s$) \\
        $s_9$ & Passive vehicle relative average speed before LC ($m/s$) \\
        $s_{10}$ & Passive vehicle standard deviation of speed before LC ($m/s$) \\
        $s_{11}$ & Passive vehicle average acceleration before LC ($m/s^2$) \\
        \hline
    \end{tabular}
    \label{tab:lane_changing_variables}
\end{table}

\textcolor{black}{The behaviors of the active and passive vehicles are jointly estimated using a quantal response equilibrium (QRE) model \cite{McKelvey1995}, which captures the interdependent nature of their decision-making while accounting for bounded rationality. Unlike the Nash equilibrium, which assumes perfectly rational agents always select the utility-maximizing option, QRE allows for probabilistic decisions based on the relative utility of each option. This probabilistic choice structure enables us to model variability in driver behavior under uncertainty or in the presence of unobserved factors. As noted by \citep{Arbis2019}, drivers’ perceptions are subject to error, leading them to behave stochastically rather than as perfect optimizers. Similarly, \citep{Wang2022} argue that drivers’ cognitive and computational limitations prevent them from fully optimizing lane-changing decisions, resulting in bounded rationality and strategic uncertainty. By tuning the rationality parameters, QRE can also incorporate the heterogeneity between AVs and HDVs, reflecting different levels of rationality across interaction types. While prior studies have applied QRE to binary driving decisions such as changing lane versus staying (for active vehicles) and yielding versus not yielding (for passive vehicles) \cite{Arbis2019, Wang2022}, our work extends this framework to model behavioral tendencies for active and passive vehicles. }

To implement the QRE approach, we first define a utility function \( U_{r, t, o} \) for each role \( r \in \{\text{active, passive}\} \), interaction type \( t \in \{\text{AV \textcolor{black}{(vs. HDV)}}, \text{HDV (vs. AV)}, \text{HDV (vs. HDV)}\} \) and outcome \( o \in \{\text{CC}, \text{CD}, \text{DC}\} \), where C and D indicate cooperative and defective behavior, respectively. 
\begin{equation}
U_{r, t, o}(\textbf{S}_i) = \bm{\beta}_{r,t,o} \cdot \tilde{\mathbf{S}_i}, 
\label{eq:utility}
\end{equation}

\noindent where \( \bm{\beta}_{r,t,o} \in \mathbb{R}^{12} \) is the coefficient vector and \( \tilde{\mathbf{S}_i} = \{1, \tilde{s}_1, \tilde{s}_2, \dots, \tilde{s}_{11}\}_i \) is the \textcolor{black}{standardized and augmented state vector for lane-changing event $i$. Note that $\tilde{\mathbf{S}_i}$ is defined by first standardizing the original state vector $\mathbf{S}_i$ and then adding a constant of 1 at the beginning to incorporate an intercept term.} In total, 216 utility parameters (2 roles × 3 interaction types × 3 outcomes × 12 coefficients) are estimated. Note that the utility of the fourth outcome, DD (both vehicles defect), is fixed to zero for all role and interaction type combinations, and serves as the baseline utility.
\begin{equation}
U_{r, t, \text{DD}}(\textbf{S}_i) = 0, \quad \forall r, t, i.
\label{eq:dd_utility}
\end{equation}

\textcolor{black}{In this formulation, the HDV is separated into two cases based on the opponent's type: HDV (vs. AV) and HDV (vs. HDV). In contrast, AV has only one case, AV (vs. HDV), due to a limitation of the WOMD dataset, which lacks AV–AV interactions. Future work could include $\text{AV (vs. AV)}$ interaction type once AV–AV lane-changing data becomes available, enabling analysis of how AV behavior varies with opponent type.}

Next, we adopt a logit-based framework in which cooperation probabilities for both active and passive vehicles are computed through an iterative process. These probabilities depend on 6 rationality parameters (2 roles × 3 interaction types), $\lambda_{r, t}$, which are estimated jointly with the utility parameters. As $\lambda_{r, t}$ increases, behavior becomes more deterministic and utility-driven, whereas $\lambda_{r, t}=0$ corresponds to purely random choice.

For the active and passive vehicles, the expected utilities associated with choosing cooperation (C) or defection (D) for a given lane-changing event $i$ are defined as:
\begin{equation}
U^C_{\text{active}, i} = p_{\text{passive}, i} \cdot U_{\text{active}, t_{a}, CC}(\textbf{S}_i) + (1 - p_{\text{passive}, i}) \cdot U_{\text{active}, t_{a}, CD}(\textbf{S}_i), 
\label{eq:utility_cooperative_active}
\end{equation}
\begin{equation}
U^D_{\text{active}, i} = p_{\text{passive}, i} \cdot U_{\text{active}, t_{a}, DC}(\textbf{S}_i) + (1 - p_{\text{passive}, i}) \cdot U_{\text{active}, t_{a}, DD}(\textbf{S}_i),
\label{eq:utility_defective_active}
\end{equation}
\begin{equation}
U^C_{\text{passive}, i} = p_{\text{active}, i} \cdot U_{\text{passive}, t_{p},CC}(\textbf{S}_i) + (1 - p_{\text{active}, i}) \cdot U_{\text{passive}, t_{p},DC}(\textbf{S}_i), 
\label{eq:utility_cooperative_defective}
\end{equation}
\begin{equation}
U^D_{\text{passive}, i} = p_{\text{active}, i} \cdot U_{\text{passive}, t_{p},CD}(\textbf{S}_i) + (1 - p_{\text{active}, i}) \cdot U_{\text{passive}, t_{p},DD}(\textbf{S}_i),
\label{eq:utility_defective_defective}
\end{equation}

\noindent where $U^C_{r, i}$ and $U^D_{r, i}$ represent the expected utilities of cooperation and defection, respectively, for the vehicle with role $r$ in a lane-changing event $i$. The term $t_a$ and $t_p$ denote the interaction type of the vehicle in active and passive roles, respectively, where $t_a, t_p \in \{\text{AV \textcolor{black}{(vs. HDV)}},\allowbreak \text{HDV (vs. AV)},\allowbreak \text{HDV (vs. HDV)}\}$. 

{
\color{black}

The probability of cooperation for a vehicle with role $r$ in a lane-changing event $i$, denoted as $p_{r,i}$, is computed using a logistic function based on the difference between the expected utilities of cooperation and defection:
\begin{equation}
p_{r, i}(\boldsymbol{\theta}) = \frac{1}{1 + \exp\left(-\lambda_{r, t_r} \cdot (U^C_{r, i} - U^D_{r, i})\right)},
\label{eq:cooperation_probability}
\end{equation}

\noindent where $\boldsymbol{\theta}$ denotes a vector of the complete set of model parameters. This vector is formed by concatenating the utility parameter vector, $\boldsymbol{\beta}\in \mathbb{R}^{216} $, and the rationality parameter vector, $\boldsymbol{\lambda}\in \mathbb{R}^{6} $. The complete parameter vector is therefore $\boldsymbol{\theta}=(\boldsymbol{\beta}, \boldsymbol{\lambda})$, which is treated as a single, flattened vector for optimization.

We perform a penalized maximum likelihood estimation to find the optimal complete parameters $\hat{\boldsymbol{\theta}}$. To prevent overfitting and encourage sparsity, we define the objective function, $J(\boldsymbol{\theta})$, as the penalized log-likelihood. This function consists of the log-likelihood summed over all observations, minus an L1 regularization term:
\begin{equation}
J(\boldsymbol{\theta}) = \left( \sum_{i} \ell(\boldsymbol{\theta}|i) \right) - \lambda_{L1} \sum_{j=1}^{216} \left| \beta_{j} \right|,
\label{eq:total_log_likelihood}
\end{equation}

\noindent where $\ell(\boldsymbol{\theta}|i)$ is the log-likelihood contribution for observation $i$, $ \lambda_{L1}$ is the regularization hyperparameter that controls the penalty strength, and $\beta_j$ are the individual coefficients within the utility parameter vector $\boldsymbol{\beta}$. The log-likelihood contribution for each observation $i$ is given by:
\begin{equation}
\begin{split}
\ell(\boldsymbol{\theta}|i) = {}&a_i \log p_{\text{active}, i}(\boldsymbol{\theta}) + (1 - a_i) \log(1 - p_{\text{active}, i}(\boldsymbol{\theta})) \\
& + b_i \log p_{\text{passive}, i}(\boldsymbol{\theta}) + (1 - b_i) \log(1 - p_{\text{passive}, i}(\boldsymbol{\theta})),
\label{eq:log_likelihood_observation}
\end{split}
\end{equation}

\noindent where \(a_i = 1\) if the active vehicle cooperates (i.e., outcome is CC or CD) and \(a_i = 0\) otherwise, indicating defection. Similarly, \(b_i = 1\) if the passive vehicle cooperates (i.e., outcome is CC or DC) and \(b_i = 0\) otherwise, indicating defection. The optimal parameter set $\hat{\boldsymbol{\theta}}$ is found by maximizing the objective function:
\begin{equation}
\hat{\boldsymbol{\theta}} = \arg\max_{\boldsymbol{\theta}} J(\boldsymbol{\theta}).
\label{eq:min_objective_function}
\end{equation}

\noindent This optimization is performed using the L-BFGS-B algorithm from SciPy, with parameter bounds set to \([-20, 20]\) for the utility coefficients and \([0.01, 10]\) for the rationality parameters.
}

\subsubsection{3.1.3. Step 3: Empirical Payoff Table Computation for Observed States}
After estimating the model parameters in Step 2, we compute the utilities for each observed lane-changing state \(\mathbf{S}_i\) using the optimized coefficients \(\hat{\boldsymbol{\beta}}_{r,t,o}\):
\begin{equation}
U_{r, t, o}(\mathbf{S}_i) = \hat{\boldsymbol{\beta}}_{r,t,o} \cdot \tilde{\mathbf{S}_i}, 
\label{eq:utility_estimated}
\end{equation}

\noindent where \(\tilde{\mathbf{S}_i}\) is the standardized \textcolor{black}{and augmented} state vector corresponding to \(\mathbf{S}_i\).

For each state \(\mathbf{S}_i\), we construct an empirical payoff table for the corresponding \textcolor{black}{observed interaction pair (i.e., active HDV vs. passive HDV; active HDV vs. passive AV; or active AV vs. passive HDV), as shown in Tables~\ref{tab:empirical_payoff_tables_combined}(a-c). This empirical payoff table allows us to analyze the interaction dynamics for each observed state \(\mathbf{S}_i\).} 

\begin{table}[tb!]
    \centering
    \caption{Empirical payoff tables for state $\mathbf{S}_i$.}
    \label{tab:empirical_payoff_tables_combined}
    \begin{tabular}{l|>{\centering\arraybackslash}m{5cm}>{\centering\arraybackslash}m{5cm}}
        
        \multicolumn{3}{l}{(a) Active HDV vs. Passive HDV} \\
        \hline
        \textbf{Active Veh.\textbackslash{} Passive Veh.} & \textbf{Cooperative} & \textbf{Defective} \\
        \hline
        \textbf{Cooperative} & \makecell{$U_{\text{active},\, \text{HDV (vs. HDV)},\, \text{CC}}(\mathbf{S}_i)$,\\ $U_{\text{passive},\, \text{HDV (vs. HDV)},\, \text{CC}}(\mathbf{S}_i)$} 
                            & \makecell{$U_{\text{active},\, \text{HDV (vs. HDV)},\, \text{CD}}(\mathbf{S}_i)$,\\ $U_{\text{passive},\, \text{HDV (vs. HDV)},\, \text{CD}}(\mathbf{S}_i)$} \\
        \rule{0pt}{3.6ex}
        \textbf{Defective} & \makecell{$U_{\text{active},\, \text{HDV (vs. HDV)},\, \text{DC}}(\mathbf{S}_i)$,\\ $U_{\text{passive},\, \text{HDV (vs. HDV)},\, \text{DC}}(\mathbf{S}_i)$} 
                           & \makecell{$0,\ 0$} \\
        \hline
        
        \multicolumn{3}{c}{} \\ 
        
        \multicolumn{3}{l}{(b) Active HDV vs. Passive AV} \\
        \hline
        \textbf{Active Veh.\textbackslash{} Passive Veh.} & \textbf{Cooperative} & \textbf{Defective} \\
        \hline
        \textbf{Cooperative} & \makecell{$U_{\text{active},\, \text{HDV (vs. AV)},\, \text{CC}}(\mathbf{S}_i)$,\\ $U_{\text{passive},\, \text{AV},\, \text{CC}}(\mathbf{S}_i)$} 
                             & \makecell{$U_{\text{active},\, \text{HDV (vs. AV)},\, \text{CD}}(\mathbf{S}_i)$,\\ $U_{\text{passive},\, \text{AV},\, \text{CD}}(\mathbf{S}_i)$} \\
        \rule{0pt}{3.6ex}
        \textbf{Defective} & \makecell{$U_{\text{active},\, \text{HDV (vs. AV)},\, \text{DC}}(\mathbf{S}_i)$,\\ $U_{\text{passive},\, \text{AV},\, \text{DC}}(\mathbf{S}_i)$} 
                           & \makecell{$0,\ 0$} \\
        \hline
        
        \multicolumn{3}{c}{} \\ 

        \multicolumn{3}{l}{(c) Active AV vs. Passive HDV} \\
        \hline
        \textbf{Active Veh.\textbackslash{} Passive Veh.} & \textbf{Cooperative} & \textbf{Defective} \\
        \hline
        \textbf{Cooperative} & \makecell{$U_{\text{active},\, \text{AV},\, \text{CC}}(\mathbf{S}_i)$,\\ $U_{\text{passive},\, \text{HDV (vs. AV)},\, \text{CC}}(\mathbf{S}_i)$} 
                             & \makecell{$U_{\text{active},\, \text{AV},\, \text{CD}}(\mathbf{S}_i)$,\\ $U_{\text{passive},\, \text{HDV (vs. AV)},\, \text{CD}}(\mathbf{S}_i)$} \\
        \rule{0pt}{3.6ex}
        \textbf{Defective} & \makecell{$U_{\text{active},\, \text{AV},\, \text{DC}}(\mathbf{S}_i)$,\\ $U_{\text{passive},\, \text{HDV (vs. AV)},\, \text{DC}}(\mathbf{S}_i)$} 
                           & \makecell{$0,\ 0$} \\
        \hline
        
        \multicolumn{3}{c}{} \\ 
        
        \multicolumn{3}{l}{\textcolor{black}{(d) Active Vehicle vs. Passive Vehicle (Generic Form)}} \\
        \hline
        \textbf{Active Veh.\textbackslash{} Passive Veh.} & \textbf{Cooperative} & \textbf{Defective} \\
        \hline
        \textbf{Cooperative} & \(R_{\text{active},t_a}(\mathbf{S}_i), R_{\text{passive},t_p}(\mathbf{S}_i)\) & \(S_{\text{active},t_a}(\mathbf{S}_i), T_{\text{passive},t_p}(\mathbf{S}_i)\) \\
        \textbf{Defective} & \(T_{\text{active},t_a}(\mathbf{S}_i), S_{\text{passive},t_p}(\mathbf{S}_i)\) & \(P_{\text{active},t_a}(\mathbf{S}_i), P_{\text{passive},t_p}(\mathbf{S}_i)\) \\
        \hline

    \end{tabular}
\end{table}

\subsubsection{3.1.4. Step 4: Identifying Social Dilemmas in Lane-Changing}
{
\color{black}
This step examines the presence of a social dilemma in each lane-changing event, represented by state $\mathbf{S}_i$. Unlike classic games, such as Prisoner's Dilemma, lane-changing involves two vehicles with distinct, interdependent roles: active and passive. To account for this asymmetry, we examine the presence of a social dilemma separately from the perspectives of active and passive vehicles.

Table~\ref{tab:empirical_payoff_tables_combined}(d) presents the interaction in a generic form to align with standard game theory conventions. In this table, $t_a$ refers to the interaction type of the active vehicle and $t_p$ refers to that of the passive vehicle. For example, if the active vehicle is an AV and the passive vehicle is an HDV, then $t_a$ is "AV (vs. HDV)" and $t_p$ is "HDV (vs. AV)". \textcolor{black}{To make the connection between our empirical utilities and this game-theoretic framework explicit, we map our calculated utilities to the four canonical payoffs for each state $\mathbf{S}_i$. First, the reward ($R$) is the payoff for mutual cooperation, defined for active and passive vehicles, respectively, as}
\begin{equation}
\label{eq:R}
R_{active, t_a}(\mathbf{S}_i) = U_{active, t_a, CC}(\mathbf{S}_i),
\end{equation}
\begin{equation}
R_{passive, t_p}(\mathbf{S}_i) = U_{passive, t_p, CC}(\mathbf{S}_i).
\end{equation}

\textcolor{black}{Next, the temptation ($T$) is the payoff earned by defecting against a cooperating opponent, defined for active and passive vehicles, respectively, as}
\begin{equation}
T_{active, t_a}(\mathbf{S}_i) = U_{active, t_a, DC}(\mathbf{S}_i),
\end{equation}
\begin{equation}
T_{passive, t_p}(\mathbf{S}_i) = U_{passive, t_p, CD}(\mathbf{S}_i).
\end{equation}

\textcolor{black}{Correspondingly, the sucker's payoff ($S$) is received by a player who cooperates while the opponent defects, defined for active and passive vehicles, respectively, as}
\begin{equation}
S_{active, t_a}(\mathbf{S}_i) = U_{active, t_a, CD}(\mathbf{S}_i),
\end{equation}
\begin{equation}
S_{passive, t_p}(\mathbf{S}_i) = U_{passive, t_p, DC}(\mathbf{S}_i).
\end{equation}

\textcolor{black}{Finally, the punishment ($P$) is the result of mutual defection, which serves as the baseline utility fixed at zero, defined for active and passive vehicles, respectively, as:}
\begin{equation}
P_{active, t_a}(\mathbf{S}_i) = U_{active, t_a, DD}(\mathbf{S}_i) = 0,
\end{equation}
\begin{equation}
\label{eq:P}
P_{passive, t_p}(\mathbf{S}_i) = U_{passive, t_p, DD}(\mathbf{S}_i) = 0.
\end{equation}

According to \citep{Macy2002}, a matrix game qualifies as a social dilemma if its payoffs satisfy the following inequalities:
\begin{itemize}
    \item $R_{r,t} > P_{r,t}$: Mutual cooperation is more favorable than mutual defection.
    \item $R_{r,t} > S_{r,t}$: Mutual cooperation is preferred over being exploited by a defector.
    \item $2R_{r,t} > T_{r,t} + S_{r,t}$: Mutual cooperation is more desirable than an equal probability of unilateral cooperation and defection.
    \item One of the following must hold:
    \begin{itemize}[label={--}]
        \item Greed ($T_{r,t} > R_{r,t}$): Exploiting a cooperator is more advantageous than mutual cooperation.
        \item Fear ($P_{r,t} > S_{r,t}$): Mutual defection is more desirable than being exploited.
    \end{itemize}
\end{itemize}
\vspace{5mm}

A matrix game that qualifies as a social dilemma can be further categorized as a `Prisoner’s Dilemma,' `Stag Hunt,' or `Chicken Game,' depending on the values of Greed ($T_{r,t} - R_{r,t}$) and Fear ($P_{r,t} - S_{r,t}$) \cite{Macy2002, Leibo2017}. Specifically:
\begin{itemize}
    \item Prisoner’s Dilemma ($T_{r,t}>R_{r,t}>P_{r,t}>S_{r,t}$): Both Greed and Fear are present.
    \item Stag Hunt ($R_{r,t}>T_{r,t}>P_{r,t}>S_{r,t}$): Fear is present, but not Greed.
    \item Chicken Game ($T_{r,t}>R_{r,t}>S_{r,t}>P_{r,t}$): Greed is present, but not Fear.
\end{itemize}
\vspace{5mm}

\textcolor{black}{It is important to note that, depending on the specific state $\mathbf{S}_i$, the interacting vehicles may or may not perceive the situation as a social dilemma. In some cases, the active vehicle may perceive the state as a social dilemma, while the passive vehicle does not. This determination is based on the computed values of $R_{r,t}(\mathbf{S}_i)$, $S_{r,t}(\mathbf{S}_i)$, $T_{r,t}(\mathbf{S}_i)$, and $P_{r,t}(\mathbf{S}_i)$ for each role \( r \in \{\text{active, passive}\} \), using the corresponding interaction types ($t=t_a$ for the active vehicle and $t=t_p$ for the passive vehicle). The aforementioned four steps are presented in
Algorithm \ref{alg:lc_framework}.
}}

\begin{algorithm}[tb!]
\caption{Framework for Identifying Social Dilemmas in Lane-Changing (LC)}
\label{alg:lc_framework}
\KwIn{Observed LC events with trajectories of active, lead, and passive vehicles}
\KwOut{Classification of observed LC states}
\BlankLine

\textbf{Step 1: Clustering Lane-Changing Behaviors} \\
\Indp
\textbullet~Extract behavioral features of active and passive vehicles (Table \ref{tab:behavioral_features_combined}) \\
\textbullet~Apply k-means clustering to classify vehicle behaviors into \textit{cooperative} or \textit{defective} strategies \\
\Indm

\textbf{Step 2: Joint Estimation of Utilities} \\
\Indp
\textbullet~Define the LC state vector $\mathbf{S}_i$ as in Table \ref{tab:lane_changing_variables} \\
\textbullet~Estimate coefficients of the utility functions using Equations \ref{eq:utility}–\ref{eq:total_log_likelihood} \\
\Indm

\textbf{Step 3: Empirical Payoff Table Computation for Observed States} \\
\Indp
\textbullet~For each observed LC state, compute the utility values to construct empirical payoff table for the observed state (Equation \ref{eq:utility_estimated}, Tables \ref{tab:empirical_payoff_tables_combined}(a-c)) \\
\Indm

\textbf{Step 4: Identifying Social Dilemmas in LC} \\
\Indp
\textbullet~Based on $R_{r,t}(\mathbf{S}_i)$, $S_{r,t}(\mathbf{S}_i)$, $T_{r,t}(\mathbf{S}_i)$, and $P_{r,t}(\mathbf{S}_i)$ for each LC state, evaluate whether the state satisfies the conditions for a social dilemma (Table \ref{tab:empirical_payoff_tables_combined}(d))\\
\Indm

\end{algorithm}

\subsection{3.2. Evolution of Cooperative Behavior in Lane-Changing}

In this subsection, we present a evolutionary game theory framework to analyze the evolution of cooperative behavior in lane-changing in a mixed traffic flow of AVs and HDVs. While various game dynamics have been proposed in the literature for evolutionary game theory \cite{Szabo1998, McAvoy2015}, our simulations fall into the category of genotypic asymmetric interactions with the pairwise comparison updating, as summarized by \citep{McAvoy2015}. They introduced asymmetric evolutionary games to generalize beyond the common symmetric framework. The specific simulation setup is presented below.

\subsubsection{3.2.1. Constructing Imputed Payoff Tables}
\textcolor{black}{A critical component of the simulation is the payoff table, which dictates the outcome of any given interaction. The evolutionary game theory framework requires a complete set of payoffs for any potential vehicle interaction. However, the WOMD dataset contains only three types of interaction pairs: active HDV vs. passive HDV, active HDV vs. passive AV, or active AV vs. passive HDV. This leaves a crucial data gap, as the simulation must also account for the unobserved active AV vs. passive AV interaction.}

\textcolor{black}{To address this, we use the utility functions, $U_{r, t, o}(\mathbf{S}_i)$, calibrated in Section 3.1.2 to construct a full set of four empirical payoff tables, one for each interaction pair, for each of the observed state $\mathbf{S}_i$. Specifically, we compute the values of $R_{r,t}(\mathbf{S}_i)$, $S_{r,t}(\mathbf{S}_i)$, $T_{r,t}(\mathbf{S}_i)$, and $P_{r,t}(\mathbf{S}_i)$, for all four interaction types \( t \in \{\text{AV (vs. HDV)}, \text{HDV (vs. AV)}, \text{HDV (vs. HDV)}\), and \(\text{AV (vs. AV)}\}\) and both roles \( r \in \{\text{active, passive}\} \), imputing unobserved cases using Equations \ref{eq:R}-\ref{eq:P}. This approach contrasts with the social dilemma analysis in Section 3.1, which used only the single payoff table corresponding to the observed interaction pair for each state. Since the utility function for the interaction type $t=$"AV (vs. AV)" could not be calibrated due to data limitations, this imputation process requires a crucial (and strong) assumption: that an AV's behavior and resulting utility in a given lane-changing state $\mathbf{S}_i$ do not depend on the type of the interacting vehicle.} It is important to note that while the aforementioned limitation in the WOMD demands such a strong assumption, upon availability of the data, the same analyses (with a complete set of payoffs) can be performed.

\textcolor{black}{This assumption allows us to generalize the calibrated utility of an AV interacting with an HDV to the unobserved active AV vs. passive AV interaction pair. Specifically, the payoffs for an active AV in this interaction are determined using its utility function when facing a passive HDV:}
\begin{equation}
U_{active, \text{AV (vs. AV)}, o}(\mathbf{S}_i):=U_{active, \text{AV (vs. HDV)}, o}(\mathbf{S}_i), 
\end{equation}

\noindent \textcolor{black}{and the payoffs for a passive AV are determined using the utility function when facing an active HDV:}
\begin{equation}
U_{passive, \text{AV (vs. AV)}, o}(\mathbf{S}_i):=U_{passive, \text{AV (vs. HDV)}, o}(\mathbf{S}_i). 
\end{equation}

\noindent \textcolor{black}{By applying this method to every state, we create a complete set of imputed payoff tables, enabling the simulation of any interaction pair. }

\subsubsection{3.2.2. Simulation Setup}
We consider a $20 \times 20$ grid, where each vehicle is positioned in a separate grid cell and adopts one of two strategies in each time step: $C$ (always cooperate) or $D$ (always defect). Additionally, each vehicle's type is either an AV or an HDV. Thus, each vehicle can be represented as a tuple $(\textit{Strategy}, \textit{Type})$. For example, a vehicle labeled $(C, AV)$ represents an AV that cooperates when acting as an active vehicle or a passive vehicle.

\textcolor{black}{Unlike traditional evolutionary games, where a single, fixed game is played repeatedly over time, our setup generalizes this process by allowing a different game to be played at each time step. In each time step, one of the observed state $\mathbf{S}_i$ is randomly selected, and each player plays the corresponding lane-changing game between itself and all of its neighbors, where neighbors are defined based on a given interaction neighbor size; for example, an interaction neighbor size of 1 means that a player interacts with any player within a Manhattan distance of 1. We assume that all players are aware of the game being played, as the game is represented by observable lane-changing state variables, as summarized in Table~\ref{tab:lane_changing_variables}. The payoff for each interaction is determined using the imputed payoff tables, with the two vehicles interacting once as an active vehicle and once as a passive vehicle to capture the average payoff across roles. The average payoff for each player at each time step is then calculated over all interactions. }

At each time step, a player $X$ updates its strategy based on the following rules. For HDVs (or AVs), strategy updates occur only when at least one neighboring HDV (or AV) is present, reflecting the assumption that drivers are more likely to learn from others of the same type. The player $X$ randomly selects a neighboring HDV (or AV), denoted as $Y$, and adopts $Y$'s strategy with probability $W$:
\begin{equation}
W = \frac{1}{1 + \exp\left[-{(E_Y - E_X)}/{K}\right]},
\label{eq:adopt_probability}
\end{equation}

\noindent where $E_X$ and $E_Y$ represent the average payoffs of players $X$ and $Y$, respectively, computed by averaging the payoffs obtained from interactions with all of their neighbors. $K$ denotes a noise parameter. $K = 0$ implies that the player $X$ will adopt $Y$'s strategy deterministically when $E_Y > E_X$. In contrast, higher values of $K$ introduce randomness into the decision-making process for choosing the behaviors.

We perform Monte Carlo simulations by varying the interaction neighbor sizes (1, 2, and 3), noise parameters (1, 2, and 3), market penetration rates (MPRs) of AVs (0.2, 0.5, and 0.8), and the frequency of social contacts (0, 2, and 4 times per 100 time steps). Social contacts are implemented by randomly shuffling the grid at a specified frequency to capture the exchange of information or experiences among drivers, reflecting that in real-world contexts, HDVs may share experiences through social networks, while AVs may exchange information via connected vehicle networks, thereby broadening their understanding of the surrounding environment \cite{Rahmati2019}. The initial proportions of cooperative and defective strategies for each vehicle type (AV or HDV) are determined from real-world observations and randomly distributed on a grid. For each combination of parameters, 20 simulations are performed to calculate the overall trends and variability of the results. This approach allows us to examine how the proportion of cooperative vehicles evolves over time under different interaction neighbor sizes, noise parameters, MPRs, and social contact frequencies. The framework for the Monte Carlo simulations is summarized in Algorithm \ref{alg:evolutionary_framework}.

\begin{algorithm}[tb!]
\caption{Framework for Analyzing Evolution of Cooperative Behavior}
\label{alg:evolutionary_framework}
\KwIn{Simulation parameters (grid size, AV market penetration rates (MPRs), interaction neighborhood sizes, noise parameter ($K$); social contact frequency, initial strategy/type distributions; $R_{r,t}(\textbf{S}_i)$, $S_{r,t}(\textbf{S}_i)$, $T_{r,t}(\textbf{S}_i)$, and $P_{r,t}(\textbf{S}_i)$ for all observed states}
\KwOut{Evolution of the cooperator ratio over time}
\BlankLine

\textbf{Step 1: Initialization} \\
\Indp
\textbullet~Construct a $20 \times 20$ grid of vehicles \\
\textbullet~Assign each grid cell a vehicle with a tuple $(\textit{Strategy}, \textit{Type})$ \\
\textbullet~Randomly assign vehicle types (AV or HDV) based on the specified MPR \\
\textbullet~Randomly initialize strategies ($C$ or $D$) based on observed dataset proportions \\
\Indm

\textbf{Step 2: Monte-Carlo Simulation with Evolution Tracking} \\
\Indp
\textbullet~For each combination of simulation parameters and scenarios, run the simulation for a fixed number of time steps (repeated 20 times for robustness) \\

\ForEach{timestep}{
    \textbullet~Shuffle grid if needed based on the social contact frequency \\
    \textbullet~Randomly select state $\textbf{S}_i$ \\
    
    \ForEach{vehicle $X$}{
    \textbullet~Identify neighbors based on the specified neighborhood size \\
    \textbullet~Simulate games between $X$ and its neighbors using $R_{r,t}(\textbf{S}_i)$, $S_{r,t}(\textbf{S}_i)$, $T_{r,t}(\textbf{S}_i)$, and $P_{r,t}(\textbf{S}_i)$ \\
    \textbullet~Compute the average payoff $E_X$ from these interactions \\
    }
    \ForEach{vehicle $X$}{
    \textbullet~Update $X$'s strategy based on the update rule \\
    }
    \textbullet~Record the proportion of cooperative vehicles 
    }
\end{algorithm}

\section{4. Results and Discussion}
In this section, we detail the results derived from analyzing the WOMD, utilizing the framework outlined previously. All analysis in this section is based on 7,636 extracted lane-changing events, which include both AVs (Waymo vehicles) and HDVs.

\subsection{4.1. Social Dilemma in Lane-Changing}
As detailed in Section 3, a four-step framework is employed to analyze the social dilemma in lane-changing. First, active and passive vehicles are categorized as either cooperative or defective. Second, utility functions are calibrated on the basis of the observed states and the corresponding behavioral choices. Third, empirical payoff table is constructed, utilizing the estimated utilities derived from the calibrated functions. Finally, the existence of a social dilemma in lane-changing is evaluated based on the empirical payoff table for each observed state.

\subsubsection{4.1.1. Step 1: Clustering Lane-Changing Behaviors}
We utilized k-means clustering with a cluster size of two to categorize vehicles in active and passive roles into two distinct groups: cooperative or defective. Table \ref{tab:kmeans_manova_combined_original}(a) presents the mean values of 10 features related to active vehicles for the two clusters identified through k-means clustering. This table also displays the results of the multivariate ANOVA (MANOVA) test, which assesses whether there are statistically significant differences between the two clusters across these 10 features. As shown in the table, the MANOVA results confirm a significant distinction between the two clusters (\textit{Wilks' lambda} = 0.3432, \textit{F} = 1459.06, \textit{p} < 0.0001), indicating that the clustering effectively differentiates the lane-changing behaviors of active vehicles into two clusters. From the descriptive statistics, the cluster 1 exhibits (1) shorter lane-changing times, (2) a higher standard deviation of speed during lane-changing, (3) greater speed gains, (4) higher maximum heading differences during lane-changing, (5) larger lane-crossing angles, (6) a greater standard deviation in yaw rate, (7) higher maximum lateral speeds during lane-changing, (8) higher maximum lateral acceleration during lane-changing, (9) a greater standard deviation of lateral acceleration during lane-changing, and (10) higher maximum acceleration during lane-changing. These characteristics suggest that cluster 1 can better represent defective lane-changing behavior, while cluster 2 can better showcase cooperative driving behavior.

\begin{table}[tb!]
    \centering
    \caption{K-means clustering and MANOVA results for (a) active and (b) passive vehicles.}
    \label{tab:kmeans_manova_combined_original}
    \begin{tabular}{lcc}
        \multicolumn{3}{l}{(a) Active Vehicles} \\
        \hline
        \textbf{Feature} & \textbf{Cluster 1} & \textbf{Cluster 2} \\
        \hline
        $x^{active}_1$: Lane-changing time (s) & 3.555 & 4.580 \\
        $x^{active}_2$: Std. of speed during LC (m/s) & 0.765 & 0.735 \\
        $x^{active}_3$: Speed gain (m/s) & 1.391 & -0.177 \\
        $x^{active}_4$: Max. heading difference during LC (rad) & 0.083 & 0.053 \\
        $x^{active}_5$: Lane crossing angle (deg) & 6.289 & 4.058 \\
        $x^{active}_6$: Std. of yaw rate during LC (rad/s) & 0.044 & 0.025 \\
        $x^{active}_7$: Max. lateral speed during LC (m/s) & 1.405 & 0.973 \\
        $x^{active}_8$: Max. lateral acceleration during LC (m/s²) & 1.470 & 0.991 \\
        $x^{active}_9$: Std. of lateral acceleration during LC (m/s²) & 0.683 & 0.449 \\
        $x^{active}_{10}$: Max. acceleration during LC (m/s²) & 1.755 & 1.288 \\
        \hline
        \multicolumn{3}{l}{\textbf{MANOVA Results}} \\
        \hline
        Wilks' lambda & \multicolumn{2}{c}{0.3432} \\
        $F$-value     & \multicolumn{2}{c}{1459.06} \\
        $p$-value     & \multicolumn{2}{c}{0.0000} \\
        \hline
        \\[2ex] 

        \multicolumn{3}{l}{(b) Passive Vehicles} \\
        \hline
        \textbf{Feature} & \textbf{Cluster 1} & \textbf{Cluster 2} \\
        \hline
        $x^{passive}_1$: Speed gain (m/s) & -5.0103 & 1.5234 \\
        $x^{passive}_2$: Max. acceleration during LC (m/s²) & 0.6556 & 1.5910 \\
        $x^{passive}_3$: Min. acceleration during LC (m/s²) & -2.9591 & -1.1956 \\
        $x^{passive}_4$: Std. of speed during LC (m/s) & 1.2662 & 0.5180 \\
        \hline
        \multicolumn{3}{l}{\textbf{MANOVA Results}} \\
        \hline
        Wilks' lambda & \multicolumn{2}{c}{0.3178} \\
        $F$-value     & \multicolumn{2}{c}{4095.14} \\
        $p$-value     & \multicolumn{2}{c}{0.0000} \\
        \hline
    \end{tabular}
\end{table}

Similarly, passive vehicles were grouped into two clusters based on four key features related to their behavior. Table \ref{tab:kmeans_manova_combined_original}(b) presents the results, where the MANOVA analysis reveals a statistically significant difference between the two clusters (\textit{Wilks' lambda} = 0.3178, \textit{F} = 4095.14, \textit{p} < 0.0001), indicating that the clustering effectively distinguishes the lane-changing behaviors of passive vehicles. Cluster 1 demonstrates characteristics typically associated with more cooperative driving: (1) lower speed gain, (2) lower maximum acceleration during lane-changing, and (3) more negative minimum acceleration during lane-changing. The increased speed variation in cluster 1 can be attributed to the need for passive vehicles to decelerate in order to create space for active vehicles to merge. Accordingly, we categorize cluster 1 as representing cooperative drivers.

\begin{figure}[tb!]
    \centering
    \includegraphics[width=\textwidth]{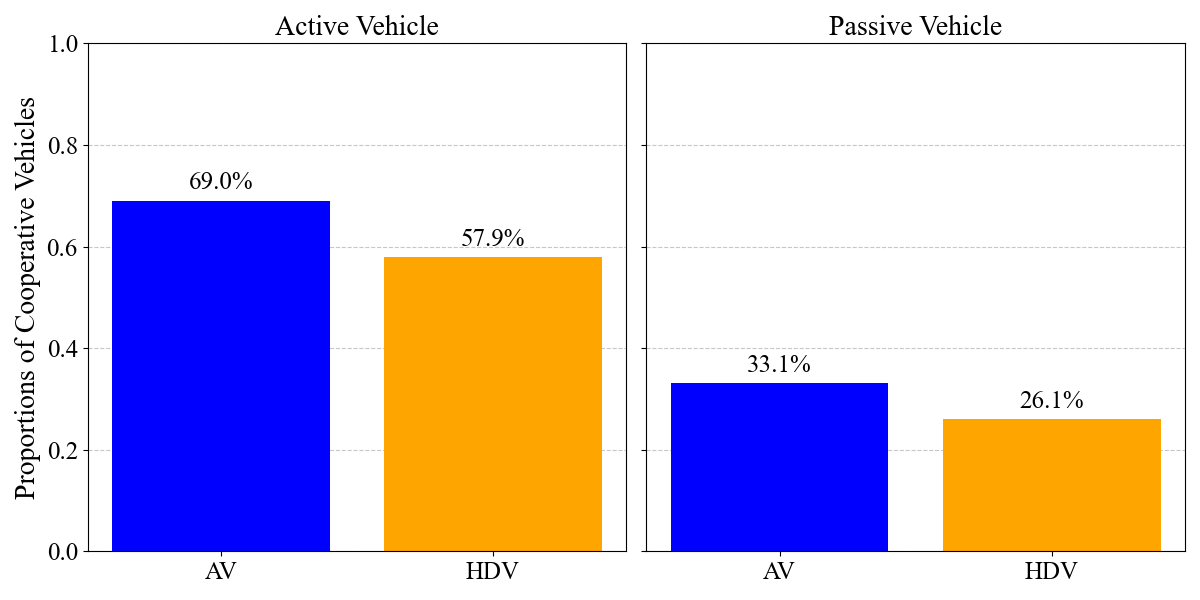}
    \caption{Proportions of cooperative AVs and HDVs across active and passive roles.}
    \label{fig:cooperative_percentage}
\end{figure}

Figure \ref{fig:cooperative_percentage} illustrates the proportions of cooperative vehicles in active and passive roles, categorized by vehicle type (AV or HDV), based on the clustering results. The graph reveals that Waymo vehicles (AVs) currently act more cooperatively than HDVs in lane-changing events, regardless of their role (active or passive). Specifically, in the active role, 69.0\% of AVs demonstrate cooperative behavior, compared to 57.9\% of HDVs. In the passive role, 33.1\% of AVs and 26.1\% of HDVs are classified as cooperative. 

Table \ref{tab:outcome_counts} summarizes the selected behaviors by interaction pairs for the active and passive vehicles across the 7,636 observed lane-changing events. As noted earlier, the WOMD dataset does not include AV-AV interactions.

\begin{table}[tb!]
    \centering
    \caption{Outcome counts by interaction pair (active vehicle vs. passive vehicle).}
    \begin{tabular}{lccccc}
        \hline
        \textbf{Outcome} & \textbf{HDV vs. HDV} & \textbf{HDV vs. AV} & \textbf{AV vs. HDV} & \textbf{AV vs. AV} & \textbf{Total} \\
        \hline
        CC & 869 & 430 & 138 & 0 & 1,437 \\
        CD & 2,200 & 617 & 230 & 0 & 3,047 \\
        DC & 507 & 134 & 35 & 0 & 676 \\
        DD & 1,821 & 525 & 130 & 0 & 2,476 \\
        \hline
        Total & 5,397 & 1,706 & 533 & 0 & 7,636 \\
        \hline
    \end{tabular}
    \label{tab:outcome_counts}
\end{table}

\subsubsection{4.1.2. Step 2: Joint Estimation of Utilities}
In this step, we developed and evaluated the performance of the QRE model for predicting the behaviors of active and passive vehicles, using the "before-lane-changing" period state variables listed in Table \ref{tab:lane_changing_variables}. Prior to modeling, we assessed the multicollinearity among the 11 normalized state variables. As detailed in Table \ref{tab:vif_data}, all the variance inflation factor (VIF) values are below the commonly accepted threshold of 5, indicating that there is no significant multicollinearity among the selected variables.

\begin{table}[tb!]
\centering
\caption{Variance inflation factors (VIF) for normalized state variables.}
\begin{tabular}{l|cccccc}
\hline
\textbf{Variable} & $\tilde{s_1}$ & $\tilde{s_2}$ & $\tilde{s_3}$ & $\tilde{s_4}$ & $\tilde{s_5}$ & $\tilde{s_6}$ \\
\hline
\textbf{VIF} & 1.509 & 1.302 & 1.676 & 1.147 & 1.789 & 1.394 \\
\hline
\end{tabular}

\vspace{3mm}

\begin{tabular}{l|cccccc}
\hline
\textbf{Variable} & $\tilde{s_7}$ & $\tilde{s_8}$ & $\tilde{s_9}$ & $\tilde{s_{10}}$ & $\tilde{s_{11}}$ & $const$ \\
\hline
\textbf{VIF} & 1.572 & 1.386 & 1.717 & 1.488 & 1.647 & 1.000 \\
\hline
\end{tabular}
\label{tab:vif_data}
\end{table}

Next, we trained the full QRE model using all available data, with the L1 regularization hyperparameter $\lambda_{L1}$ set as 0.1. The full model achieved a log-likelihood of -7,521.2, higher than the null model's -10,542.4, that includes only intercept terms for each outcome. The likelihood ratio test statistic of 6,042.4 with 204 degrees of freedom ($p<0.0001$) indicates that the full model significantly outperforms the null model. McFadden’s pseudo-$R^2$ of 0.287 further demonstrate the model's improved explanatory power over the baseline.

The estimated rationality parameters, $\lambda_{r,t}$, of the full model capture the degree of rationality in decision-making for each role, with higher values indicating more deterministic behavior (see Table~\ref{tab:lambda_params}). Overall, passive vehicles exhibit higher levels of rationality than active vehicles, \textcolor{black}{consistent across all interaction types. Specifically, the average $\lambda$ for passive roles (2.13) exceeds that for active roles (1.53), suggesting that passive vehicles make more consistent and less stochastic decisions.} Moreover, AVs generally show more deterministic behavior than HDVs in both active and passive roles. \textcolor{black}{For instance, when interacting with passive HDVs, the rationality of active AVs ($\lambda=1.82$) surpasses that of active HDVs ($\lambda=1.23$); similarly, when interacting with active HDVs, passive AVs ($\lambda=2.18$) exhibit higher rationality than passive HDVs ($\lambda=1.87$).} Furthermore, HDVs behave more deterministically when interacting with AV than when interacting with other HDV, in both roles. The lowest $\lambda$ values are observed in HDV (vs. HDV) for both \textcolor{black}{active ($\lambda=1.23$) and passive ($\lambda=1.87$) roles}, indicating greater variability likely driven by the heterogeneity of human behaviors, in contrast to the more deterministic patterns exhibited by AVs.

\begin{table}[tb!]
    \centering
    \caption{Estimated rationality parameters (\(\lambda_{r,t}\)) for the full QRE model.}
    \begin{tabular}{lcc}
        \hline
        \textbf{Role} ($r$) & \textbf{Interaction Type} ($t$) & $\bm{\lambda_{r,t}}$ \\
        \hline
        Active &  AV \textcolor{black}{(vs. HDV)}& 1.8228 \\
        Active & HDV (vs. AV) & 1.5426 \\
        Active & HDV (vs. HDV) & 1.2267 \\
        Passive & AV \textcolor{black}{(vs. HDV)}& 2.1848 \\
        Passive & HDV (vs. AV) & 2.3476 \\
        Passive & HDV (vs. HDV) & 1.8667 \\
        \hline
    \end{tabular}
    \label{tab:lambda_params}
\end{table}

To further validate the model, we performed cross-validation of the QRE model using 10 random stratified 70/30 train-test splits, ensuring that the proportion of interaction pairs (active HDV vs. passive HDV, active HDV vs. passive AV, active AV vs. passive HDV) was preserved in each split. 

As shown in Table \ref{tab:QRE_cooperation_predict}, the QRE model demonstrated a strong predictive performance in capturing both the mean and variance in interactive decisions during lane-changing maneuvers. Following \citep{Arbis2019}, the expected number in cooperating decisions for each role $r$ was estimated as $\sum_i p_{r,i}$, and the corresponding standard deviation as $\sqrt{\sum_i p_{r,i} (1-p_{r,i})}$. For active vehicle, the predicted expected cooperation count (1,341.88) closely matched the observed average (1,347.40). Similarly, for passive vehicle, the predicted expected cooperation number (632.88) aligned well with the observed value (630.60). In terms of variability, the model slightly underestimated the standard deviation for active vehicle (22.40 predicted vs. 23.56 observed), and for passive vehicle (16.14 vs. 21.37). Overall, these results suggest that the QRE framework effectively captures aggregate decision patterns under uncertainty, acknowledging that drivers' perceptions include errors.

\begin{table}[tb!]
\centering
\caption{Comparison of observed cooperation decisions with QRE. Results are averaged over 10 test sets. For each set, there are 2,291 interactions.}

\begin{tabular}{l|cc|cc}
\hline
\multirow{2}{*}{\textbf{Equilibrium}} 
& \multicolumn{2}{c|}{\textbf{Active Veh. Cooperation Decisions}} 
& \multicolumn{2}{c}{\textbf{Passiv Veh. Cooperation Decisions}} \\
& \textbf{Average Expected} & \textbf{Average Stdev} 
& \textbf{Average Expected} & \textbf{Average Stdev} \\
\hline
Observed      & 1,347.40 & 23.56 & 630.60 & 21.37 \\
QRE & 1,341.88 & 22.40 & 632.88 & 16.14 \\
\hline
\end{tabular}
\label{tab:QRE_cooperation_predict}
\end{table}

\textcolor{black}{The comparison between observed and QRE-predicted cooperation decisions, however, does not evaluate prediction accuracy at the individual interaction level. Instead, it examines whether the model captures aggregate behavioral patterns, such as the average number and variability of cooperative decisions across multiple interactions. To further assess the model's ability to predict specific outcomes, }we evaluated its performance on joint outcomes (CC, CD, DC, and DD) using precision, recall, and F1-score, as presented in Table~\ref{tab:confusion_metrics}. These metrics are calculated based on the standard confusion matrix definitions:
\begin{table}[tb!]
    \centering
    \caption{Precision, recall, and F1-score for each outcome, averaged over 10 test sets. \textcolor{black}{Values in parentheses indicate the standard deviation across splits.}}
    \begin{tabular}{lccc}
        \hline
        \textbf{Outcome} & \textbf{Precision} & \textbf{Recall} & \textbf{F1-Score} \\
        \hline
        CC & 0.5562 (0.0203) & 0.5172 (0.0193) & 0.5356 (0.0134) \\
        CD & 0.5188 (0.0062) & 0.7315 (0.0215) & 0.6069 (0.0067) \\
        DC & 0.4652 (0.0377) & 0.2742 (0.0317) & 0.3442 (0.0329) \\
        DD & 0.5773 (0.0217) & 0.3717 (0.0136) & 0.4519 (0.0120) \\
        \hline
        Macro Average & 0.5294 (0.0426) & 0.4736 (0.1722) & 0.4846 (0.0979) \\
        \hline
    \end{tabular}
    \label{tab:confusion_metrics}
\end{table}
\begin{equation}
\text{Precision} = \frac{\text{TP}}{\text{TP} + \text{FP}},
\label{eq:precision}
\end{equation}
\begin{equation}
\text{Recall} = \frac{\text{TP}}{\text{TP} + \text{FN}},
\label{eq:recall}
\end{equation}
\begin{equation}
\text{F1-Score} = \frac{2 \cdot \text{Precision} \cdot \text{Recall}}{\text{Precision} + \text{Recall}},
\label{eq:f1}
\end{equation}

\noindent where TP, FP, and FN denote true positives, false positives, and false negatives, respectively. Macro-averaged scores across all outcomes are 0.5294 for precision, 0.4736 for recall, and 0.4846 for F1-score, indicating overall moderate predictive ability across outcome types. This is reasonable given that behaviors are classified as cooperative or defective, which are not perfectly separable. In contrast, previous studies often modeled more clearly defined binary tasks, such as lane change versus no lane change or give way versus do not give way \cite{Talebpour2015, Arbis2019, Wang2022}, which are comparatively simpler to distinguish than the behaviors examined in this paper.

\subsubsection{4.1.3. Step 3: Empirical Payoff Table Computation for Observed States}
For all 7,636 extracted lane-changing events, we calculated the state variables $s_1, \dots, s_{11}$, resulting in 7,636 state vectors ($\mathbf{S}_1, \mathbf{S}_2, \dots, \mathbf{S}_{7636}$) from the dataset. \textcolor{black}{Using the calibrated coefficients, we derived an empirical payoff table for each state $\mathbf{S}_i$ for the corresponding interaction pair. This process yielded one of the three empirical payoff tables (Tables \ref{tab:empirical_payoff_tables_combined}(a-c)) for each state.}

\subsubsection{4.1.4. Step 4: Identifying Social Dilemmas in Lane-Changing}

In the final step, the $R_{r,t},\ S_{r,t},\ T_{r,t}$, and $P_{r,t}$, as presented in Table \ref{tab:empirical_payoff_tables_combined}(d), are used to examine the presence of social dilemmas in lane-changing. 

\begin{figure}[tb!]
    \centering
    \includegraphics[width=\textwidth]{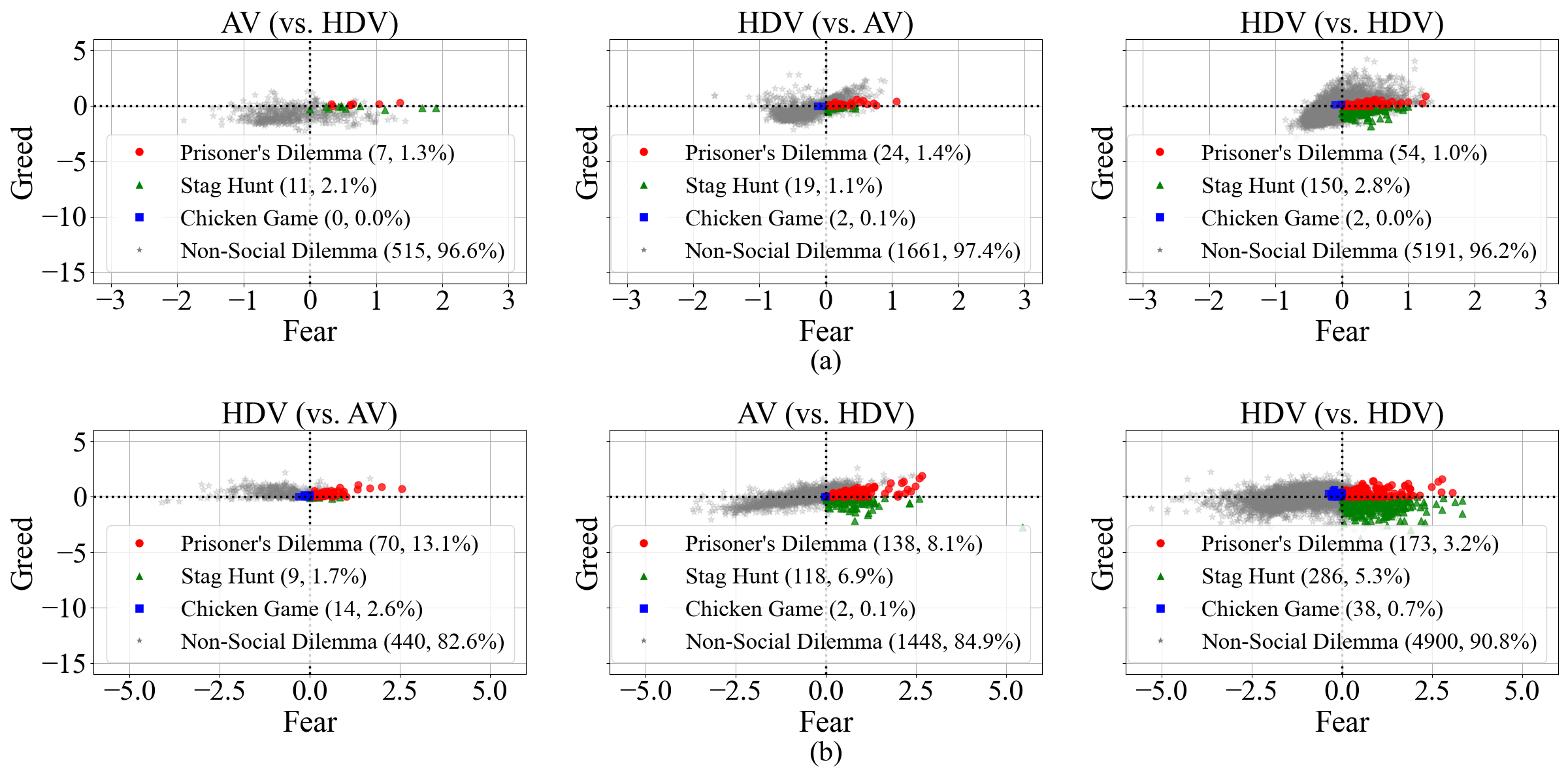}
    \caption{\textcolor{black}{Social dilemmas in lane-changing based on empirical payoffs from the perspective of (a) active and (b) passive vehicles, for each interaction type.}}
    \label{fig:social_dilemma}
\end{figure}

Figure~\ref{fig:social_dilemma} displays scatter plots of Fear (defined as $P - S$) and Greed (defined as $T - R$) on the $x$ and $y$ axes, respectively, for each interaction type, from the perspectives of active (Figure~\ref{fig:social_dilemma}(a)) and passive (Figure~\ref{fig:social_dilemma}(b)) vehicles. \textcolor{black}{Each plot visualizes a number of points corresponding to the number of observed states $\mathbf{S}_i$ for each interaction type, as previously summarized in Table \ref{tab:outcome_counts}. Each point represents how a vehicle in role $r$ and interaction type $t$ perceives the game structure of the lane-changing interaction at a given state $\mathbf{S}_i$.}

The analysis confirms the presence of social dilemmas in lane-changing interactions. \textcolor{black}{Out of the observed 7,636 states, active vehicles perceive approximately 4\% of the states as social dilemmas, while passive vehicles perceive around 11\%.} For both roles, most social dilemmas are classified as Prisoner's Dilemma or Stag Hunt, while Chicken Game is rarely observed across all interaction types.

Tables \ref{tab:social_dilemma_combined}(a-c) provide a detailed breakdown of how various states are categorized into different game types for all observed interaction pairs. The results show that even within the same state, the interacting vehicles may perceive the game structure differently. This implies that the diversity of game structures in lane-changing, including the presence of multiple types of social dilemma, may limit the potential of AVs to inherently promote greater cooperation, despite their initial higher tendency to cooperate, as shown in Figure \ref{fig:cooperative_percentage}. Therefore, to explore whether cooperation emerges in mixed traffic within these intricate lane-changing interactions, we apply evolutionary game theory in the following subsection.

\begin{table}[tb!]
    \centering
    \caption{\textcolor{black}{Social dilemma comparison for the three interaction pairs.}}
    \label{tab:social_dilemma_combined}
    \begin{tabular}{l *{5}{>{\raggedleft\arraybackslash}p{1.9cm}}}
        \multicolumn{6}{l}{(a) Active HDV vs. Passive HDV} \\
        \hline
        \multicolumn{1}{c}{} & \multicolumn{5}{c}{\textbf{Passive HDV}} \\
        \cline{2-6}
        \textbf{Active HDV} & \textbf{PD} & \textbf{SH} & \textbf{CG} & \textbf{Non-SD} & \textbf{Total} \\
        \hline
        \textbf{PD}     & 1   & 9   & 0   & 44   & 54 \\
        \textbf{SH}     & 6   & 17  & 0   & 127  & 150 \\
        \textbf{CG}     & 0   & 0   & 0   & 2    & 2 \\
        \textbf{Non-SD} & 166 & 260 & 38  & 4,727 & 5,191 \\
        \hline
        \textbf{Total}  & 173 & 286 & 38  & 4,900 & 5,397 \\
        \hline
        \\
        \multicolumn{6}{l}{(b) Active HDV vs. Passive AV} \\
        \hline
        \multicolumn{1}{c}{} & \multicolumn{5}{c}{\textbf{Passive AV}} \\
        \cline{2-6}
        \textbf{Active HDV} & \textbf{PD} & \textbf{SH} & \textbf{CG} & \textbf{Non-SD} & \textbf{Total} \\
        \hline
        \textbf{PD}     & 0   & 0   & 0   & 24   & 24 \\
        \textbf{SH}     & 0   & 0   & 0   & 19   & 19 \\
        \textbf{CG}     & 0   & 0   & 0   & 2    & 2 \\
        \textbf{Non-SD} & 138 & 118 & 2   & 1,403 & 1,661 \\
        \hline
        \textbf{Total}  & 138 & 118 & 2   & 1,448 & 1,706 \\
        \hline
        \\
        \multicolumn{6}{l}{(c) Active AV vs. Passive HDV} \\
        \hline
        \multicolumn{1}{c}{} & \multicolumn{5}{c}{\textbf{Passive HDV}} \\
        \cline{2-6}
        \textbf{Active AV} & \textbf{PD} & \textbf{SH} & \textbf{CG} & \textbf{Non-SD} & \textbf{Total} \\
        \hline
        \textbf{PD}     & 0   & 0   & 0   & 7    & 7 \\
        \textbf{SH}     & 0   & 0   & 0   & 11   & 11 \\
        \textbf{CG}     & 0   & 0   & 0   & 0    & 0 \\
        \textbf{Non-SD} & 70  & 9   & 14  & 422  & 515 \\
        \hline
        \textbf{Total}  & 70  & 9   & 14  & 440  & 533 \\
        \hline
        \multicolumn{6}{l}{\footnotesize \textit{Note}: PD: Prisoner's Dilemma, SH: Stag Hunt, CG: Chicken Game, Non-SD: Non-Social Dilemma}
    \end{tabular}
\end{table}

\subsection{4.2. Evolution of Cooperative Behavior in Lane-Changing}

\textcolor{black}{
To investigate how cooperative behavior evolves in repeated lane-changing interactions, we adopt an evolutionary game theory framework detailed in Section 3.2. In our Monte Carlo simulations, one of the 7,636 observed states is randomly selected at each time step, and every vehicle plays the corresponding game with its neighbors. Each vehicle plays both the active and passive roles once with all neighbors and updates its strategy based on the average payoff, calculated using the imputed payoff tables constructed for that state as described in Section 3.2.1. }

\textcolor{black}{Tables \ref{tab:empirical_payoff_tables_combined_s1}(a-d) illustrate the imputed payoff tables for an arbitrarily chosen state, $\mathbf{S}_1$. As a reminder, we compute the imputed payoff tables for all four possible interaction pairs for the simulation, although the true interaction pair for state $\mathbf{S}_1$ was active HDV vs. passive HDV. Because of our assumption that the estimated coefficients for an AV's behavior when interacting with an HDV can be generalized to its behavior when interacting with another AV, an active (or passive) AV receives the same payoff regardless of the interacting vehicle’s type.} For instance, when an active AV interacts with a passive HDV, the payoff to the active AV under mutual cooperation is –0.239 (Table \ref{tab:empirical_payoff_tables_combined_s1}(c)), identical to the payoff when the active AV interacts with a passive AV (Table \ref{tab:empirical_payoff_tables_combined_s1}(d)). 

Importantly, even within the same lane-changing state, the empirical payoffs can vary substantially across interaction pairs. For example, when paired with a defective passive HDV, an active HDV earns a higher payoff by defecting ($0 > -0.140$, Table \ref{tab:empirical_payoff_tables_combined_s1}(a)), whereas with a defective passive AV, the same active HDV earns a higher payoff by cooperating ($0.116 > 0$, Table \ref{tab:empirical_payoff_tables_combined_s1}(b)). These intricate payoff structures motivate the use of empirical game theory to analyze the evolution of cooperative and defective behaviors in mixed traffic during repeated lane-changing events. The initial proportions of cooperative AVs and HDVs in all Monte Carlo simulations are set to 51\% and 42\%, respectively, based on Figure \ref{fig:cooperative_percentage}.

\begin{table}[tb!]
    \centering
    \caption{Imputed payoff tables for an arbitrarily chosen state $\mathbf{S}_1$.}
    \label{tab:empirical_payoff_tables_combined_s1}
    \renewcommand{\arraystretch}{1.2}
    \begin{tabular}{l|>{\centering\arraybackslash}m{3.5cm}>{\centering\arraybackslash}m{3.5cm}}
        
        \multicolumn{3}{l}{(a) Active HDV vs. Passive HDV} \\
        \hline
        \textbf{Active Veh.\textbackslash{} Passive Veh.} & \textbf{Cooperative} & \textbf{Defective} \\
        \hline
        \textbf{Cooperative} & \makecell{-0.030, -1.963} & \makecell{-0.140, -1.939} \\
        \textbf{Defective} & \makecell{-0.433, 2.077} & \makecell{0, 0} \\
        \hline
        \multicolumn{3}{c}{} \\[-1.5ex]

        \multicolumn{3}{l}{(b) Active HDV vs. Passive AV} \\
        \hline
        \textbf{Active Veh.\textbackslash{} Passive Veh.} & \textbf{Cooperative} & \textbf{Defective} \\
        \hline
        \textbf{Cooperative} & \makecell{0.719, -1.040} & \makecell{0.116, -1.821} \\
        \textbf{Defective} & \makecell{0.196, 1.867} & \makecell{0, 0} \\
        \hline
        \multicolumn{3}{c}{} \\[-1.5ex]

        \multicolumn{3}{l}{(c) Active AV vs. Passive HDV} \\
        \hline
        \textbf{Active Veh.\textbackslash{} Passive Veh.} & \textbf{Cooperative} & \textbf{Defective} \\
        \hline
        \textbf{Cooperative} & \makecell{-0.239, -1.446} & \makecell{0.203, -1.314} \\
        \textbf{Defective} & \makecell{0.153, 1.319} & \makecell{0, 0} \\
        \hline
        \multicolumn{3}{c}{} \\[-1.5ex]

        \multicolumn{3}{l}{(d) Active AV vs. Passive AV} \\
        \hline
        \textbf{Active Veh.\textbackslash{} Passive Veh.} & \textbf{Cooperative} & \textbf{Defective} \\
        \hline
        \textbf{Cooperative} & \makecell{-0.239, -1.040} & \makecell{0.203, -1.821} \\
        \textbf{Defective} & \makecell{0.153, 1.867} & \makecell{0, 0} \\
        \hline

    \end{tabular}
\end{table}

\begin{figure}[tb!]
    \centering
    \includegraphics[width=\textwidth]{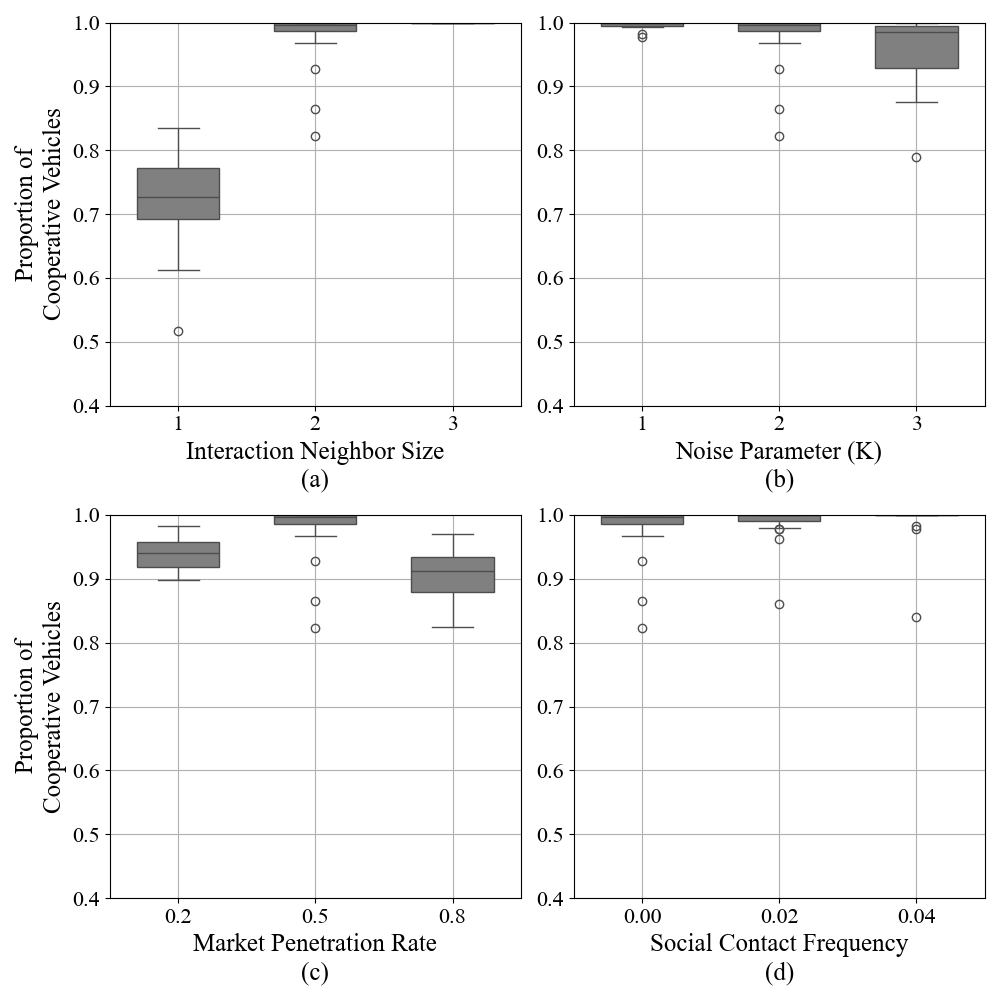}
    \caption{Results of Monte Carlo simulations after 200 time steps: Sensitivity to (a) interaction neighbor size, (b) noise parameter, (c) market penetration rate, and (d) social contact frequency.}
    \label{fig:EGT_sensitivity}
\end{figure}

Figures \ref{fig:EGT_sensitivity}(a-d) show the sensitivity analysis for (a) the interaction neighbor size, (b) the noise parameter ($K$), (c) the MPR of AVs, and (d) the social contact frequency, with each panel showing boxplots of the proportion of cooperative vehicles after 200 time steps. Unless otherwise stated, the default values used in the figures are an interaction neighbor size of 2, a noise parameter of 2, an MPR of 0.5, and a social contact frequency of 0. Each parameter combination was simulated 20 times, and the proportion of cooperative vehicles represents the share of cooperative vehicles within the entire vehicle population in mixed traffic.

As depicted in Figure \ref{fig:EGT_sensitivity}(a), increasing the number of neighbors each vehicle interacts with significantly improves the level of cooperation. When the interaction neighbor size is set to 3, all 20 simulation runs converge to a fully cooperative population. This trend suggests that broader interactions enable vehicles to better recognize cooperative behavior, thereby accelerating the evolution toward mutual cooperation in mixed traffic.

Additionally, as shown in Figure \ref{fig:EGT_sensitivity}(b), the cooperation declines with higher noise parameter. With a lower noise parameter ($K=1$), players make more deterministic decisions based on payoff differences, promoting consistent cooperative behavior. In contrast, a higher noise parameter ($K=3$) introduces greater randomness in strategy updates regardless of payoff disparity, as governed by Equation \ref{eq:adopt_probability}. To balance realism and behavioral convergence, we set the default values for noise parameter to 2, as previously noted.

Figure \ref{fig:EGT_sensitivity}(c) compares cooperation levels under three different AV MPR scenarios. Interestingly, the MPR of 0.5 achieves the highest proportion of cooperative vehicles. This outcome can be attributed to a well-mixed population that facilitates enough interactions between the same vehicle types, which are essential for the strategy updates in the genotypic asymmetric interactions with the pairwise comparison updating setting \cite{McAvoy2015}. In contrast, more homogeneous populations limit such interactions for the minority type, reducing opportunities for cooperation to emerge.

Finally, Figure \ref{fig:EGT_sensitivity}(d) shows that the proportion of cooperative vehicles increases with more frequent social contacts. A social contact frequency of 0.04 corresponds to four random shuffling of the grid during 100 time steps, meaning the grid was shuffled every 25 steps. The sensitivity results for social contact frequency are consistent with those for interaction neighbor size, as both indicate that more interaction with diverse vehicles fosters more cooperative behavior. These results indicate that facilitating information exchange through HDV social networks or AV connected vehicle networks may help foster cooperative behavior.

\begin{figure}[tb!]
    \centering
    \includegraphics[width=\textwidth]{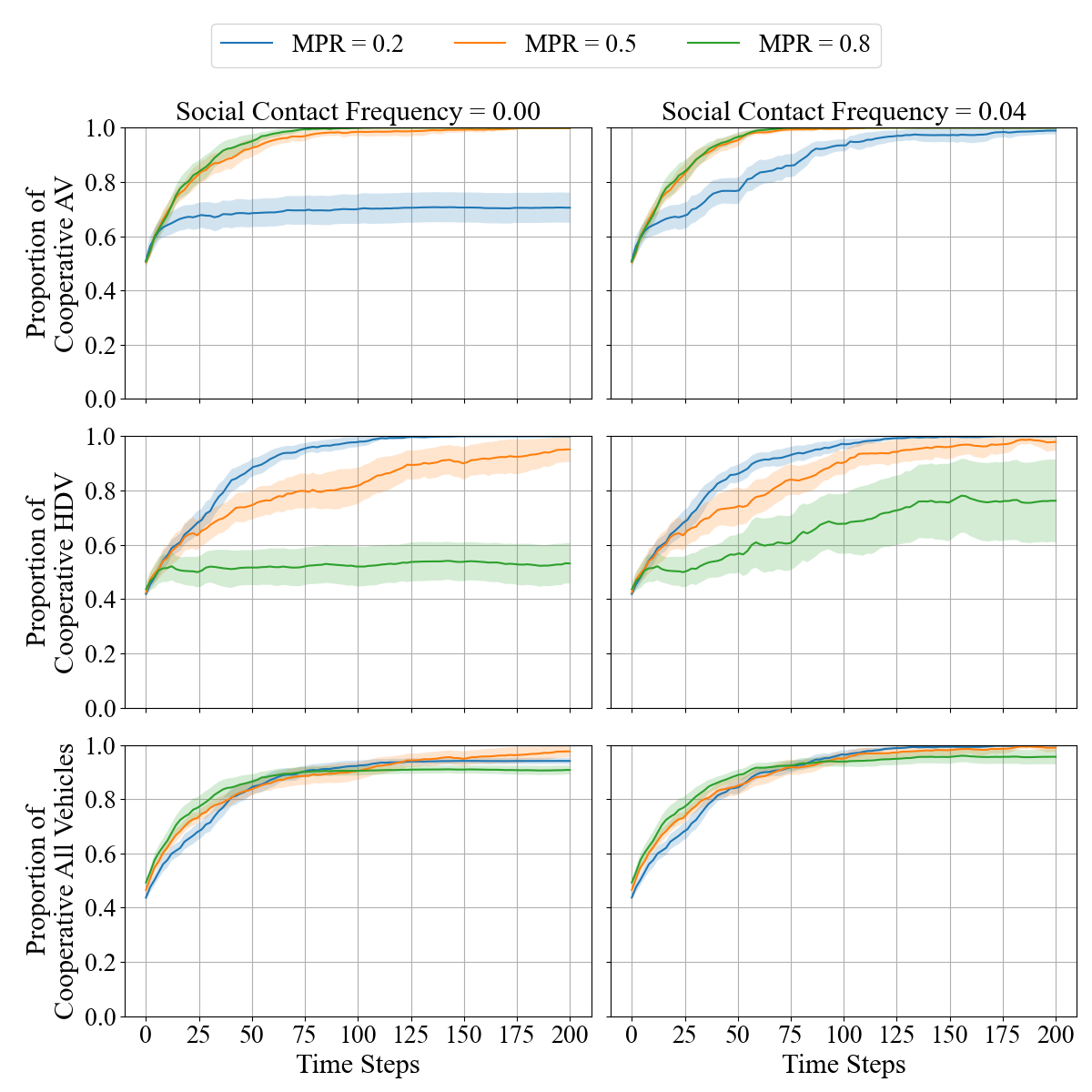}
    \caption{Results of Monte Carlo simulations: Proportion of cooperative AVs (top), HDVs (middle), and all vehicles (bottom) over time. The shaded regions represent the 95\% confidence intervals across 20 simulation runs.}
    \label{fig:EGT_av_hdv}
\end{figure}

We further evaluate the evolution of the proportion of cooperative vehicles over time for three different MPRs, separately for AVs (top), HDVs (middle), and all vehicles (bottom) in Figure \ref{fig:EGT_av_hdv}. The left panels depict trends with a social contact frequency of 0, while the right panels show that of 0.04. The solid line represents the mean proportion of cooperative vehicles across 20 Monte Carlo simulations, while the shaded regions represent the 95\% confidence interval.

First, with no social contacts, a clear trend emerges among AVs (top left). Their cooperation rate increases as their presence on the road grows. When AVs dominate (MPR=0.8), nearly all AVs eventually adopt cooperative behavior in lane-changing interactions. In contrast, when AVs are a minority (MPR=0.2), their cooperation levels remain relatively stable over time, showing little deviation from their initial proportion. A similar pattern of increasing cooperation is observed among HDVs across all MPRs (middle left), although the extent of this increase varies. In HDV-dominated scenarios (MPR=0.2), their cooperation rate rises more significantly over time. As the proportion of AVs increases, the growth in HDV cooperation becomes more modest, potentially due to fewer interactions with other HDVs, which may limit the reinforcement of cooperative behavior through social learning within their own group.

With social contacts, the overall trend of increasing cooperation over time in all MPRs remains consistent. However, some differences are notable. Even without formal statistical testing, the graph suggests that the proportion of cooperative AVs (top) grows more quickly with social contacts in HDV-dominated scenarios (MPR=0.2). Likewise, the proportion of cooperative HDVs (middle) increases faster with social contacts in AV-dominated scenarios (MPR=0.8). These patterns indicate that social contacts, through information sharing, can help promote cooperation particularly among minority vehicle types.

Importantly, despite the presence of a social dilemma in some lane-changing events, all simulated scenarios exhibit a consistent increase in cooperation over time for both AVs and HDVs. In other words, empirical payoff tables estimated from real-world trajectory data suggest that repeated lane-changing interactions promote cooperative behavior for both AVs and HDVs, regardless of the AV penetration rate in mixed traffic.

\section{5. Conclusions}
This paper presents a data-driven framework for evaluating cooperative and defective behaviors in lane-changing among AVs and HDVs, using real-world trajectory data from the WOMD. The framework first classifies the behaviors of active and passive vehicles as cooperative or defective using k-means clustering, then quantifies their utilities to construct empirical payoff tables. These tables are subsequently used to identify the presence of social dilemmas in real-world lane changes. Finally, we apply evolutionary game theory to examine how cooperative behavior evolves under varying AV MPRs.

Our analysis reveals that AVs currently exhibit a higher proportion of cooperative behaviors than HDVs in both active and passive roles during lane changes. \textcolor{black}{Additionally, approximately 4\% of the observed lane-changing states are classified as social dilemmas for active vehicles, and 11\% for passive vehicles. Most of these dilemmas fall into the categories of Stag Hunt or Prisoner's Dilemma, with Chicken Game occurring only rarely.} Even with the presence of social dilemmas in lane-changing, our Monte Carlo simulations indicate that repeated lane-changing interactions tend to increase cooperative behavior in mixed traffic over time not only in AV- or HDV-dominated traffic, but also in well-mixed environments. To the best of our knowledge, this study offers the first empirical evidence of social dilemmas emerging in lane-changing behavior within mixed traffic, and of the evolution of cooperative behavior during such interactions.

\textcolor{black}{One limitation of this study is the lack of AV-AV interaction data, which prevents analysis of how AV behavior may differ when interacting with another AV compared to an HDV.} Future research could explore this when such data becomes available. Furthermore, while this study focused only on the behaviors during successful lane changes, it would be valuable to analyze real-world trajectory data that include failed lane-changing attempts due to unyielding passive vehicles. Doing so would offer a more comprehensive understanding of lane-changing dynamics in mixed traffic. Another research direction will be to explore the possibility of incorporating the proposed framework into game-theoretical lane-changing or duration models.

\section{Acknowledgements}
This work was supported by the National Science Foundation under Grant No. 2047937.

\section{Declaration of generative AI and AI-assisted technologies in the writing process}
The authors acknowledge the use of AI-assisted tools (such as ChatGPT) for language editing and grammar refinement during manuscript preparation. No AI tool was used for generating novel content, data analysis, or drawing conclusions. All responsibility for the accuracy and integrity of the manuscript remains with the authors.

\section{Declaration of Conflicting Interests}
The authors declared no potential conflicts of interest with respect to the research, authorship, and/or publication of this article.

\newpage

\bibliographystyle{partc}
\bibliography{Draft}
\end{document}